\documentclass[usenatbib]{mnras}

\usepackage{newtxtext,newtxmath}

\usepackage[T1]{fontenc}
\usepackage{ae,aecompl}

\usepackage{graphicx}	
\usepackage{amsmath}	
\usepackage{booktabs}
\usepackage{booktabs}
\usepackage[table,xcdraw]{xcolor}
\usepackage{soul}
\usepackage{float}
\usepackage{ulem}
\usepackage{color}           
\usepackage{chngcntr}
\usepackage{txfonts}
\usepackage{subcaption}
\usepackage{lscape}
\usepackage{longtable}
\usepackage{newtxtext,newtxmath}
\usepackage{caption}
\usepackage{multicol} 

\title[Non-stationarity in quasi-periodic pulsations]{Prevalence of non-stationarity in quasi-periodic pulsations (QPPs) associated with M- and X- class solar flares}

\author[T. Mehta et al.]{
T. Mehta,$^{1}$\thanks{E-mail: t.mehta.1@warwick.ac.uk}
A.-M. Broomhall,$^{1,2}$
and  L. A. Hayes, $^{3}$ 
\\
$^{1}$ Centre for Fusion, Space and Astrophysics, Department of Physics, University of Warwick, Coventry, CV4 7AL, UK\\
$^{2}$ Centre for Exoplanets and Habitability, University of Warwick, Coventry CV4 7AL, UK \\
$^{3}$  European Space Agency (ESA), European Space Research and Technology Centre (ESTEC), Keplerlaan 1, 2201 AZ Noordwijk, The Netherlands}

\date{Accepted 2023 May 26. Received 2023 May 26; in original form 2022 Oct 18}

\pubyear{2023}

\begin{document}
\label{firstpage}
\pagerange{\pageref{firstpage}--10}
\maketitle

\begin{abstract}
Quasi-periodic pulsations (QPPs) are frequently observed in solar and stellar flare emission, with recent studies suggesting that an increasing instantaneous period is a common characteristic of QPPs. Determining the prevalence of non-stationarity in QPPs contributes to a better understanding of which mechanism(s) is (are) responsible in QPP generation. We obtain the rate of period evolution from QPPs in 98 M- and X- class flares from Solar Cycle~24 with average periods between 8-130~s and investigate the prevalence of QPP non-stationarity. We also investigate whether the presence of a Coronal Mass Ejection (CME) impacts the period evolution of QPPs. We analyse soft X-ray lightcurves obtained from GOES’ X-Ray Sensor (XRS) and assess the dominant periods in the impulsive and decay phases of the flares using the Fast Fourier Transform. We relate the rate of period evolution to flare duration, peak flare energy, and average QPP period. We find evidence of non-stationarity in 81$\%$ of the flares assessed, with most QPPs exhibiting a period evolution of $\leq$10~s between the impulsive and decay phases, of which 66$\%$ exhibited an apparent period growth and 14$\%$ showed an apparent period shrinkage. We find a positive correlation between the absolute magnitude of period evolution and the duration of the flare and no correlation between the period evolution of the QPPs and flare energy or CME presence. Furthermore, we conclude that non-stationarity is common in solar QPPs and must be accounted for in flare analysis.

\end{abstract}



\section{Introduction}
 
The emission from a solar flare often demonstrates fluctuations in intensity as a function of time. These fluctuations are known as Quasi-Periodic Pulsations (QPPs) and are characterised as repetitive bursts with similar time-scales that can range from seconds to several tens of seconds \citep[][]{nak09, vandoorsselarer2016, 2020Kupriyanova}. QPPs are identified across the entire electromagnetic spectrum of flare emissions, meaning that they are typically a multi-wavelength phenomenon \citep[e.g. see][]{clarke2021}. While non-thermal hard X-ray and microwave observations clearly demonstrate the most prominent pulsations during a flare, measurements from the past Solar Cycle with Sun-as-a-star soft X-ray (SXR) and extreme ultraviolet (EUV) observations have shown that small-amplitude QPPs are a very common feature of solar flares \citep[][]{simoes2015soft, 2018Dominique, Hayes_2020}.
 
The study of solar flare emission fluctuations extends beyond our solar system as stellar flare QPPs have been extensively observed \citep{2000Zhilyaev, 2016Pugh, broomhall2019}. These QPPs observed in stellar flares are largely similar in characteristics to those observed in solar QPPs, which strengthens the case for a solar-stellar analogy for QPPs \citep[see][for an overview on recent advances in observations of stellar QPPs]{2021Zimovets}. Therefore a better understanding of the mechanism(s) driving QPPs in solar flares is likely to lead to advances in stellar QPPs.  

The question as to what causes these repetitive flare emissions has been the topic of significant discussion \citep[][]{mclaughlin2018}, with over fourteen different mechanisms suggested to date \citep[See][and references therein for an overview on generation mechanisms]{2020Kupriyanova, 2021Zimovets}. The proposed generation mechanisms can be sorted into three groups; \textit{1.} Mechanisms that modulate the direct release of plasma emissions as the result of MHD oscillations; \textit{2.} Mechanisms where MHD waves modulate the efficiency of energy release; \textit{3.} Mechanisms based on spontaneous quasi-periodic energy release. Despite the growing number of mechanisms proposed to underpin the generation of QPPs, we are not yet in a position to confidently identify which mechanism is responsible and it seems likely that there are multiple mechanisms at play in generating QPPs. 

There is an expanding catalogue of QPPs which exhibit non-stationary properties, with both the phase, period, and amplitude varying in time \citep[see][for review]{nak2019}. For example, period drifts have been identified in several flares \citep{kupriyanova2010, simoes2015soft, 2018Kolotkov}, with it common to find the decay phase periods longer than the associated impulsive phase periods \citep[e.g.][]{hayes2016quasi, Hayes_2020}. Notably, in some cases, QPPs can be observed to extend late into the decay phase of solar flares and illustrate systematic increases in periods \citep{dennis2017detection, 2019Hayes}. There is a growing need to understand how the periods evolve over flares, whether period drifts are a common feature of flare QPPs, and whether the period drifts are systematic based on flare class, duration, or whether they are eruptive or not. We also need to address the prevalence of non-stationarity in solar QPPs, as the majority of detection methods used currently rely on a periodogram-based approach. As discussed in \citet[][]{broomhall2019b}, periodogram-based approaches tend to be less successful when detecting a non-stationary QPP. It is likely that we are missing, or at best, poorly characterising, the presence and behaviour of many QPPs by assuming their dominant periods are stationary. In quantifying the proportion of QPPs that exhibit non-stationarity we can better discern which analysis methods are the most appropriate to use when searching and categorising QPPs.

In this work we explore the nature of QPP period drifts by investigating whether non-stationarity is an inherent feature of QPPs. To achieve this we build upon the work of \cite{Hayes_2020} and we present a comparison of the dominant periods (the periodicity that corresponds to the largest peak relative to the confidence level in a power spectrum) in the impulsive phase of the flare (characterised as the time from the start of the flare to the time corresponding to flare maximum) and the decay phase (after the flare peak) in QPPs from M- and X- class flares from Solar Cycle 24. By examining the prevalence of QPPs that show evidence of non-stationarity we can potentially classify the different types of QPPs present in solar flare emission, and help constrain which mechanisms can drive QPPs.

\section{Observations and Analysis Methods} \label{Methods}

\subsection{Data}
\label{subsec:data}

To select a list of flares for which to perform this study, we utilise a list of M- and X- GOES class flares from 1$\textsuperscript{st}$ February 2011 to 31$\textsuperscript{st}$ December 2018 (i.e. Solar Cycle 24) that demonstrated strong evidence of QPP signatures in their emission from the study of \cite{Hayes_2020}. This list consists of 205 flare events that showed enhanced Fourier power in the periodograms of the GOES-XRS 1--8~\AA\ channel observations. We further analyse this list of flares by focusing on the same 1--8~\AA\ channel from the GOES-15 satellite which has a cadence of 2.047~s, and focus on analysing the impulsive and decay phases of the flares independently to identify features of non-stationarity and period drift.

To determine the duration of the impulsive phase, we use the flare start and peak times defined within the GOES flare catalogue produced by the National Oceanic and Atmospheric Administration (NOAA). NOAA define the flare start time as the first minute in a sequence of four minutes wherein there is a steep monotonic increase in the 1--8~\AA\ channel and the final flux value is greater than the first by a factor of 1.4. The flare peak time is the time at which the flares soft X-ray emission reaches its flare peak energy, which is its maximal value as measured in the  1--8~\AA\, channel. For our analysis, we limit the time window of the decay phase to the same duration as the impulsive phase. We use this method of choosing the end times rather than using the end times defined within the GOES flare catalogue. This is because of our implementation of  criteria (\textit{ii}) (discussed in Section~\ref{subsec:method}) which requires 5 or more full cycles in each phase of the flare. This means that for a flare with impulsive/ decay phases of unequal length, each phase has a different upper limit on the maximal periodicity that can be obtained. This discrepancy in the upper limit threatens to artificially induce artefacts in the data. Therefore for consistency we limit the time window of the decay phase to the same duration as the impulsive phase, as can be seen in Fig.~\ref{fig:flare40}. However for most events the end times we chose and those defined by the GOES catalogue were similar.

To examine whether the presence of a Coronal Mass Ejection (CME) correlates with the appearance or magnitude of a period evolution of the QPPs, we use the publicly available SOHO/LASCO CME catalogue, to determine which flares had associated CMEs.

\subsection{Method}
\label{subsec:method}

We separate the flare into the impulsive and decay phases, we perform a Fast Fourier Transform (FFT) on each phase and test whether a periodic signature is present above a 95$\%$ confidence level. We obtain the confidence levels by making use of the technique outlined in \citet{Pugh_2017} which is based upon the work in \citet{Vaughan}. This method involves fitting the power spectrum with a broken power law, which accounts for both the presence of red and white noise in the signal and avoids the problems that can arise in assessing the significance of an identified periodic signature when detrending data. Using this fitting we determined the 95$\%$ confidence level. Any peaks in the power spectra above these confidence levels were deemed to be statistically significant. 

We make use of this method as it was determined to be highly effective in robustly detecting the period of QPPs in a Hare and Hound exercise \citep[see Table 5 in][]{broomhall2019}. However we note that periodogram-based methods do fail in the detection of non-stationary QPPs, \citep[as discussed in Section 5.4 of][]{broomhall2019}, whereas EMD and other methods that allow for varying time scales were more effective in detecting these QPPs. We chose not to use EMD as it struggles with non-detrended data and can be a user intensive process. Instead we opted to use the Fourier based method on a windowed signal. This constrained our study to periodicities that are relatively stationary within their shortened durations. This is a clear limitation in our work as we are unlikely to detect periodicities that evolve rapidly in either flare phase due to spectral leakage in the resulting power spectra. In theory, we may be able to detect some of the more rapidly evolving periodicities in the data using shorter or overlapping windows, should they exist, however preliminary studies showed that reducing the duration of the signals resulted in fewer overall detections which we attribute to the decreased number of oscillatory cycles in the data. The flare database that this study uses originates from a periodogram-based approach \citep[][]{Hayes_2020}, and so we find it likely that the FFT will produce statistically meaningful results in both phases. This technique allows for a statistically sound analysis that can be applied to a large sample of flares. 

We note that recent literature suggests that the significance of peaks in periodograms can be overestimated for non-stationary QPPs if segments are poorly selected. We follow a suggested mitigation strategy put forward in \citet{Hubner2022} by splitting the flare event into two phases and only assessing events in which there is similar statistically significant QPP-like behaviour in both segments, as outlined below. 

After performing a FFT on both phases of a given flare and obtaining the dominant periods, we discard the data if it does not fulfil the following criteria; \textit{(i)} the periods obtained for both phases must be statistically significant above a 95$\%$ confidence level, \textit{(ii)} the periods for both phases must be less than one tenth of the full duration of the flare, \textit{(iii)} the periods for both phases must be greater than four times the cadence of the data (i.e. both periods must be greater that 8.19~s), and \textit{(iv)} the impulsive phase period must not be greater or smaller than the decay phase period by more than a factor of eight. Criteria (\textit{ii}) aims at targeting QPPs with at least five full oscillatory cycles in both the impulsive and decay phase. We also restrict our periods to be greater than four times the cadence of the dataset (criteria \textit{(iii)}). This is because we believe detections of periods smaller than this are unreliable when detected by GOES alone and must be accompanied by other data sources with better time resolution. Finally we believe QPPs that exhibit a change in period by a factor larger than eight (criteria (iv)) implies that the QPP in the impulsive phase does not correspond to the QPP in the decay phase. This could, for example, be caused by two periodicities present in the signal but one not reaching the 95$\%$ confidence level due to a change in the signal to noise ratio. It is important to state that the absence of the above criteria being met for a given flare event does not necessarily imply that no QPPs were present. Rather there may have been QPPs that were not statistically significant in both phases, or one whose period evolution was outside of the criteria we put forward. However we restrict our study to these criteria in the interest of reliability and consistency of results. This resulted in 98 flares which fulfilled all the criteria, which are discussed in Section~\ref{sec:results}.

We define the term \textit{period drift} to measure the change in period from the impulsive phase to that in the decay phase, equal to Period$_{Decay}$ - Period$_{Impulsive}$. A positive period drift implies an increase in dominant period from the impulsive phase to the decay phase and vice versa. We emphasise that there may be multiple processes present in generating the QPPs and a positive period drift does not imply the growth in period of a singular QPP process- for example such an effect could similarly be produced by a process producing shorter period QPPs decaying in amplitude in tandem with a secondary longer period process growing in amplitude. This would result in a growth in dominant period across the two phases, i.e. a positive period drift. 

We determine the average period of the flare by taking the mean of the dominant periods in the impulsive and decay phases. As we are examining the prevalence of non-stationarity in QPPs we avoid taking an FFT of the entire duration of the flare to obtain the average period, as a non-stationary signal that has significant period evolution is not well suited to the FFT which assumes a stationary input. It is possible that a non-stationary signal which evolves over several frequencies will show evidence of spectral leakage in its associated power spectrum, leading to any dominant peaks being smeared out, and presenting no statistically significant peaks. This is naturally still an issue to be considered when assessing only the impulsive or decay phase and any quickly evolving periodicity is likely to be obscured in the same manner, which may lead to a number of false negatives in our results when statistically significant periods are not found in our analysis. However by splitting the flare into sections we still should be able to observe some periods with sufficiently slow evolution and still pick up on their long-term non-stationarity. 

We determine the errors on the periods from the impulsive and decay phases by use of the standard approach, and propagate these errors to obtain the errors on period drift and the average period \citep[See Section 4.2.1 in ][for a detailed discussion on error propagation]{hughes2010measurements}.

\begin{figure}
\centering
\includegraphics[width=\linewidth]{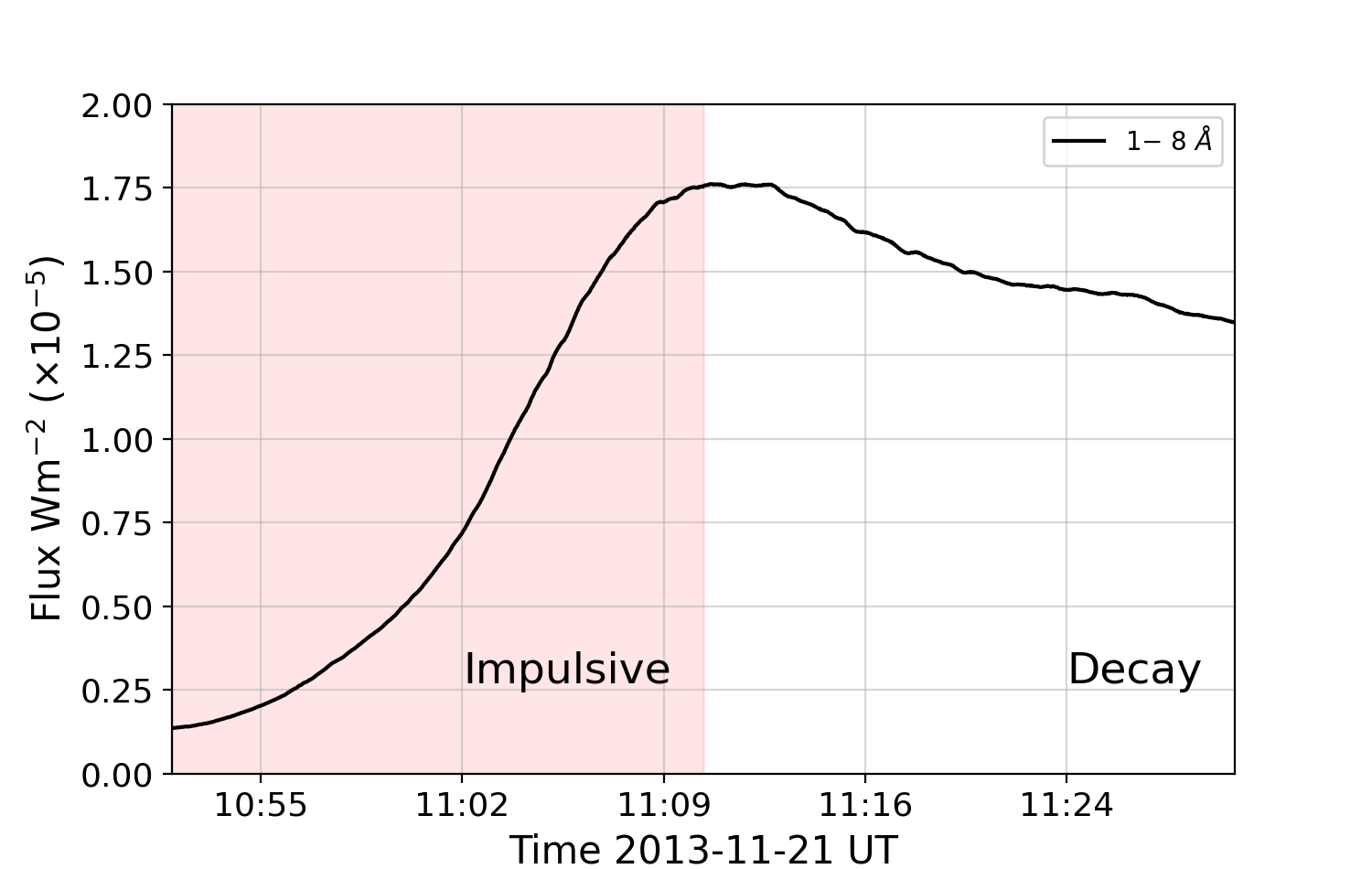}
\caption{Profile of Flare~40 in GOES-XRS 1--8~\AA, where the impulsive phase is shaded in red and the decay phase is unshaded. The analysed impulsive and decay phases are equal in duration and are delineated by flare maximum which occurs at approximately 11:10 UT. }
\label{fig:flare40}
\end{figure}

\begin{figure}
\begin{subfigure}[b]{0.45\textwidth}
\centering
\includegraphics[width=.85\linewidth]{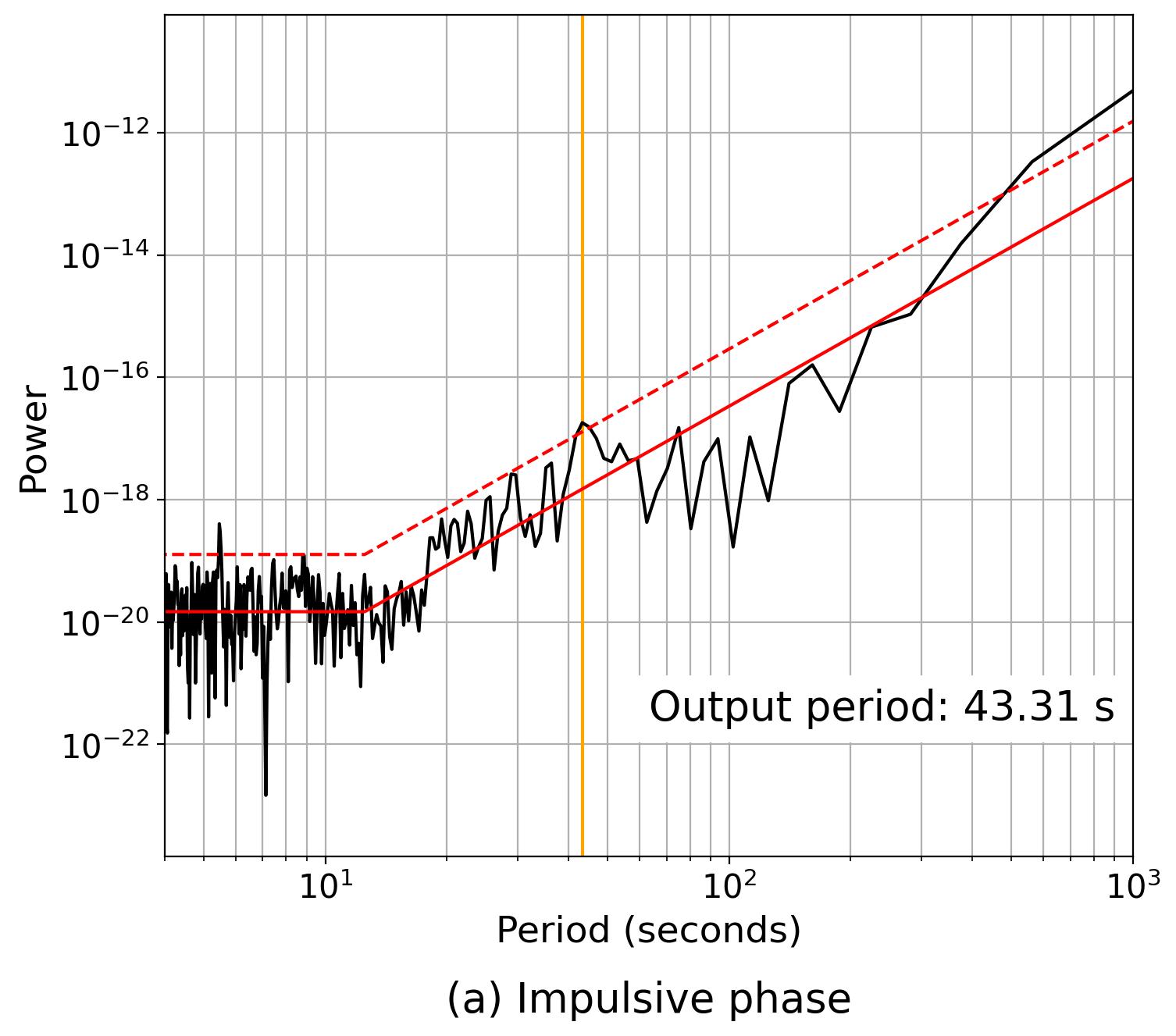}
\caption*{}
\label{fig:impulse}
\end{subfigure}
\begin{subfigure}[b]{0.45\textwidth}
\centering
\includegraphics[width=.85\linewidth]{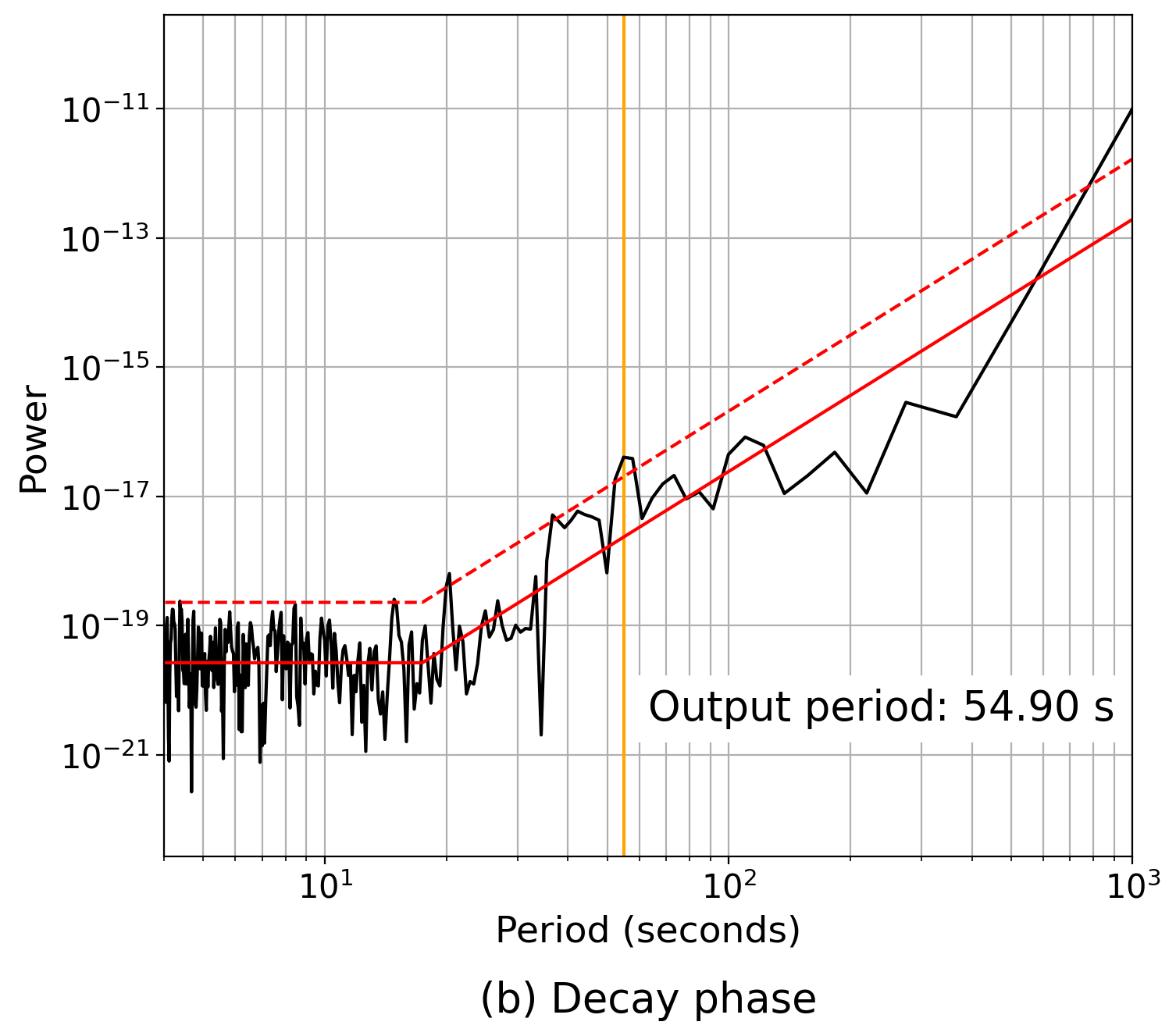}
\caption*{}
\label{fig:decay}
\end{subfigure}
\caption{Fourier spectra of Flare~40. \textit{Top}: Fourier spectrum of the impulsive phase. \textit{Lower}: Fourier spectrum of the decay phase. Fits of the spectra by broken power laws are shown by solid red lines, and the 95$\%$ confidence levels are indicated with dashed red lines. Statistically significant peaks (indicated by vertical orange lines) can be seen corresponding to periods of 43.3~s in the impulsive phase and 54.9~s in the decay phase. \label{fig:impulseanddecay}}
\end{figure}

Fig~\ref{fig:flare40} shows the 1--8~\AA\ lightcurve for Flare~40  where the duration of the flare has been symmetrically split into the impulsive phase until flare maximum, and the decay phase. Fig.~\ref{fig:impulseanddecay} shows the Fourier spectra of Flare~40's impulsive and decay phases, which show significant periods of 43.3$\genfrac{}{}{0pt}{2}{+1.7}{-1.6}$ and 54.9$\genfrac{}{}{0pt}{2}{+2.8}{-2.5}$~s respectively, corresponding to a period drift of 11.6$\genfrac{}{}{0pt}{2}{+3.3}{-3.0}$~s.

\section{Results}
\label{sec:results}

We examine 205 solar flares from M- and X-class flares over Solar Cycle~24, resulting in 98 flares that show statistically significant periods in both the impulsive and decay phases of the flare that have both periods greater than four times the cadence of the dataset, less than one tenth of the full duration of the flare, and separated in period by no more than a factor of eight. We consider a period drift to be statistically significant if its absolute magnitude is greater than 4.09~s, which is twice the cadence of the data. This is a cautious approach as we see that the errors on periods are generally smaller than the cadence. Of these 98 flares, 19 (equivalent to 19$\%$) showed no significant period drift. Of the remaining 79 QPPs, 65 (66$\%$ of the sample) exhibited a positive period drift where the dominant period appears to increase from the impulsive to the decay phase. 14 flares (14$\%$) exhibited a negative period drift where the dominant period appears to shrink between the phases.

\begin{figure}
\centering
\includegraphics[width=\linewidth]{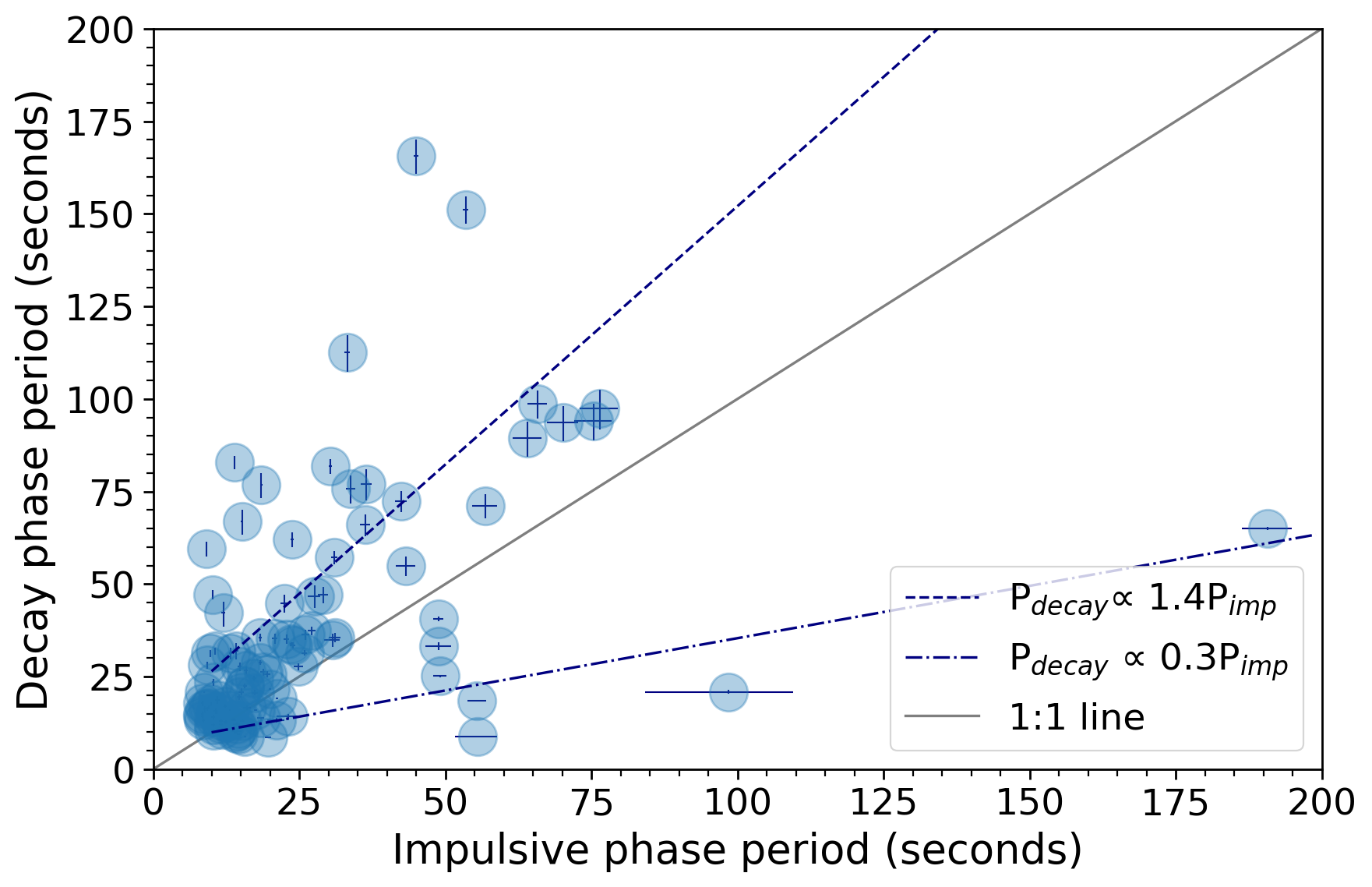}
\caption{QPP impulsive phase periods against decay phase periods. A 1:1 ratio line (which indicates no period drift) is shown as a solid black line. The impulsive phase periods are between 8 -- 75~s, and are approximately similar across the all flares, whereas the decay phase periods have a larger spread between 8 -- 110~s. The line of best fit for QPP periods that grew between the impulsive and decay phases is shown as a dashed blue line, and the line of best fit for period that shrunk is shown as a dot dashed blue line. This figure uses new data to recreate Fig.~10 from \citet{Hayes_2020}.\label{fig:imp_dec}}
\end{figure}

Fig.~\ref{fig:imp_dec} shows the relationship between the impulsive and decay phase periods of the 98 flares examined. It can be seen that the majority of results appear above the 1:1 ratio line shown in solid black, which indicates more QPPs have a larger decay phase period than impulsive phase period. For the QPPs showing an apparent period growth, the decay phase periods are loosely correlated to the impulsive phase periods by a factor of approximately $\sim$1.4, although there is significant scatter for events with decay phase periods greater than 40~s. This correlation agrees well with the factor of $\sim$1.6 that was found in a similar analysis, shown in Fig.~10 of \cite{Hayes_2020}, which shows the difference in periods detected during the impulsive and decay phases of 28 flaring events (20 of which overlap with the study presented in this paper). We note that the authors found that 26 of these events (92$\%$) showed a larger decay phase period than impulsive phase period and their factor is based on the fitting of all 28 events, not just those that show period growth. For the 65 QPPs exhibiting positive period drift, the median period drift is 13$\genfrac{}{}{0pt}{2}{+13}{-6}$~s where the errors correspond to the periods in the upper and lower 25$\textsuperscript{th}$ percentile. Similarly the median negative period drift for the 14 flaring events is -10$\genfrac{}{}{0pt}{2}{+3}{-24}$~s.

\begin{figure}
\centering
\includegraphics[width=\linewidth]{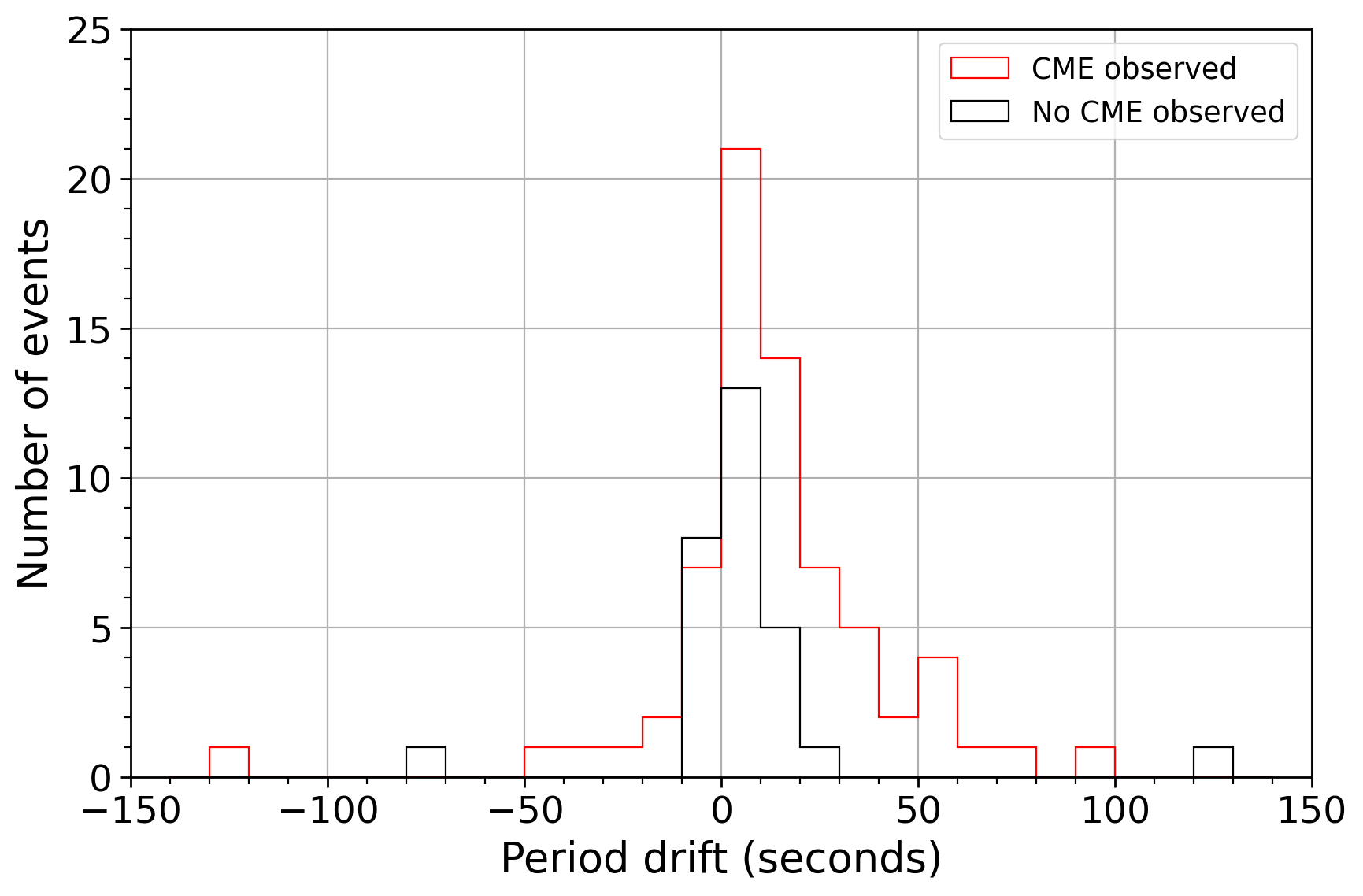}
\caption{ Histogram of period drifts of QPPs, separated by CME association. The QPPs seen in flares associated with CMEs are given in red, and those not associated with a CME are shown in black.}
\label{fig:pd_hist_cme}
\end{figure}

We examine whether the presence of a CME associated with the flare impacts the distribution of period drifts in QPPs. Of the 98 QPPs, 69 were associated with a CME and 29 were not. Fig.~\ref{fig:pd_hist_cme} shows the histogram of period drifts in QPPs from flares associated with CMEs (red) and those from flares not associated with CMEs (black). The distributions of the two sets are reasonably similar with median period drifts of 10$\genfrac{}{}{0pt}{2}{+13}{-9}$~s for the CME associated flares and 5$\genfrac{}{}{0pt}{2}{+4}{-6}$ for the non-CME associated flares. The maximal and minimal period drifts across both groups are also similar with the CME-associated group having maximal and minimal period drifts of 98 and -126~s, and the non-CME associated group with 121 and -76~s.

\begin{figure}
\centering
\includegraphics[width=.97\linewidth]{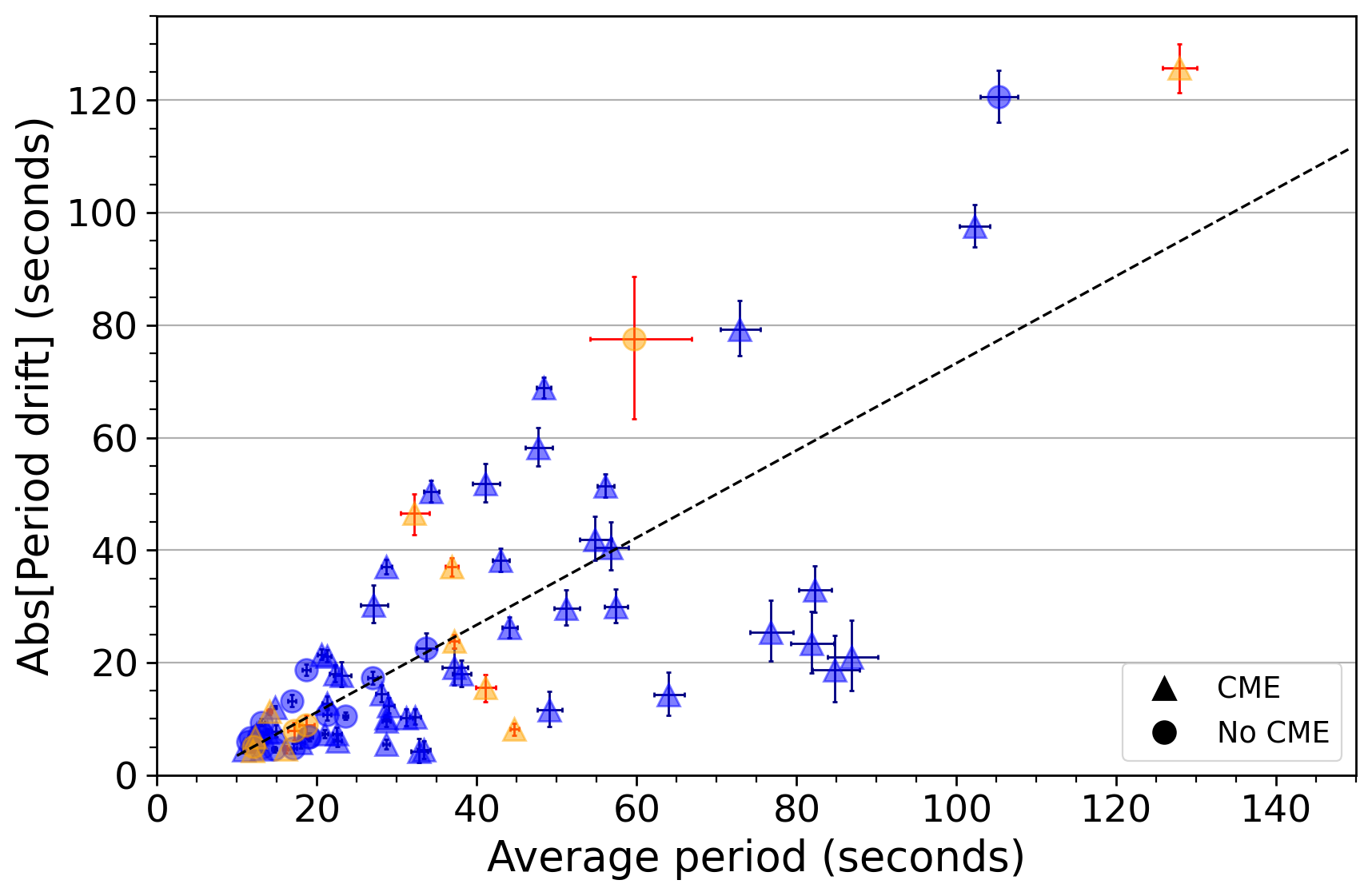}
\caption{ Average QPP period plotted against the absolute magnitude of the QPP period drift. Positive period drifts, indicating a growth in dominant period, are shown in blue, and negative period drifts are shown in orange. QPPs from flares associated with CMEs are indicated by a triangle marker whereas those not associated with QPPs are shown with bullet points. The Pearson correlation coefficient is 0.76 indicating a positive correlation. A linear fit of the data is shown as a black dashed line.}
\label{fig:pd_per}
\end{figure}

Fig.~\ref{fig:pd_per} shows the relationship between absolute period drift and average QPP period. Positive period drifts are shown in blue, and the absolute values of negative period drifts are shown in orange. QPPs associated with a CME are shown with a triangle and non-CME associated events are marked with a circle. The meanings of the colours and symbols used in Fig.~\ref{fig:pd_per} are consistent for the remainder of this paper. A positive correlation, with a Pearson correlation coefficient of 0.76, can be seen between the average period of the QPPs and the magnitude of the period drift. However we emphasise that this artificial correlation is largely induced by the selection criterion (iv) of the flares.

\begin{figure}
\centering
\includegraphics[width=\linewidth]{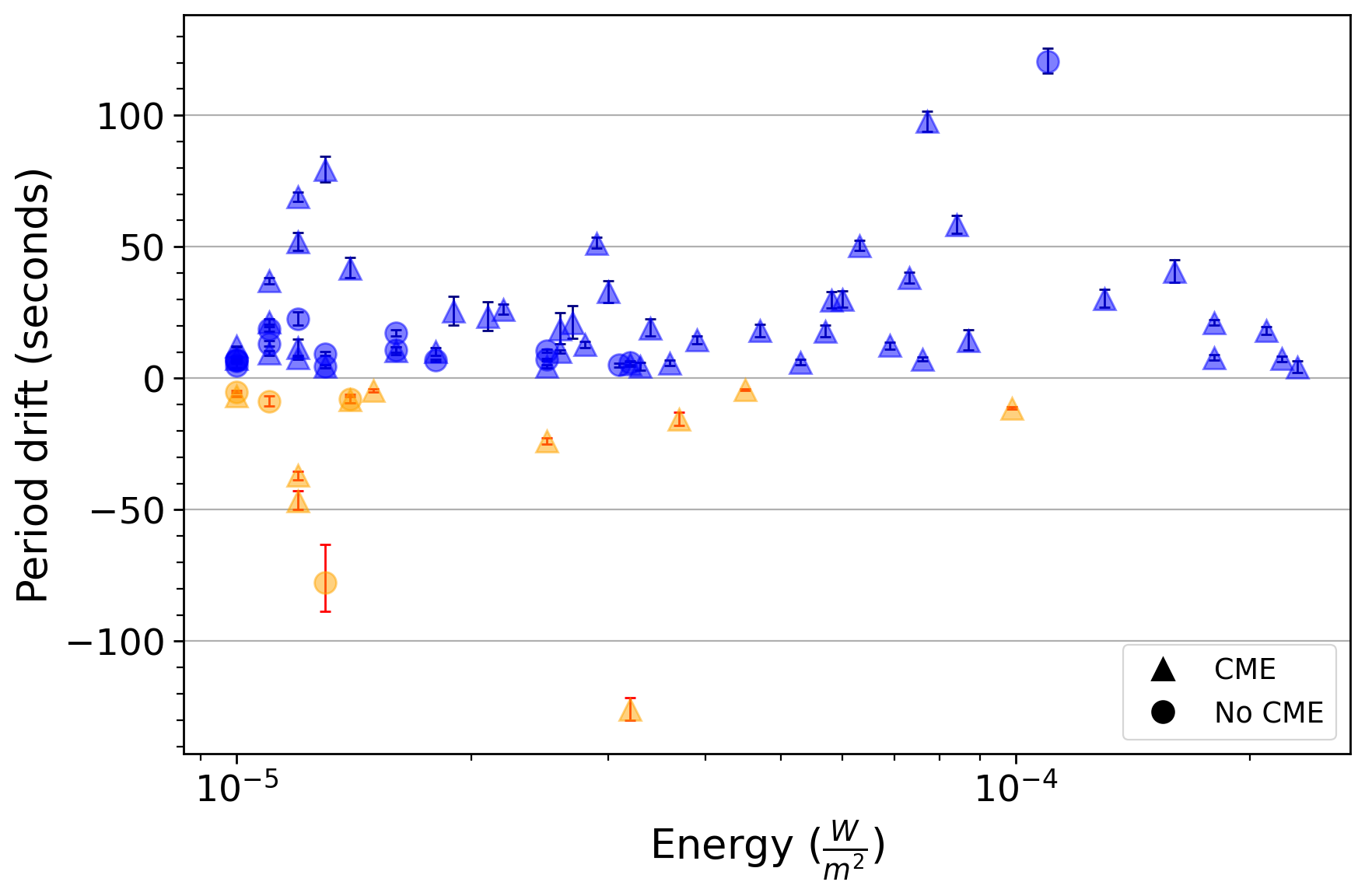}
\caption{Peak flare energy as measured in \textit{GOES} 1 -- 8 \AA~ plotted against QPP period drift with no correlation. The meanings of colours and symbols are as given in Fig.~\ref{fig:pd_per}. }
\label{fig:pd_en}
\end{figure}

Maximal flare energy, which is taken to be the maximal emission as measured in the 1--8 \AA\, channel, and QPP period drift are seen to have no correlation in Fig.~\ref{fig:pd_en}. As expected the flares not associated with CMEs are more commonly found at lower energies but this distinction has no significant effect on the magnitude or direction of the period drifts observed. 

\begin{figure}
\centering
\includegraphics[width=\linewidth]{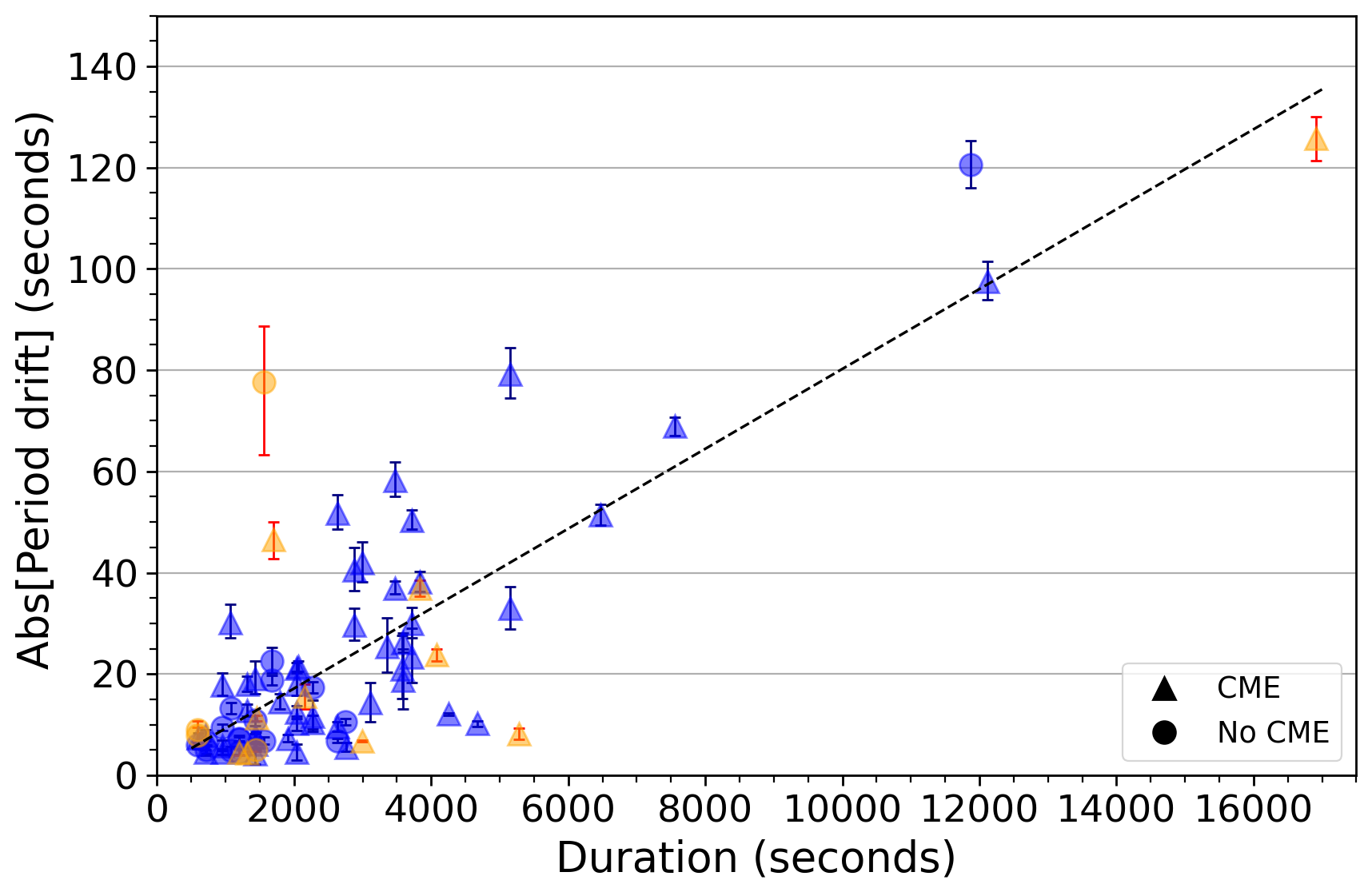}
\caption{Flare duration plotted against the absolute magnitude of the QPP period drift. The Pearson correlation coefficient is 0.82, indicating a positive correlation. A linear fit of the data is shown as a black dashed line. The meanings of colours and symbols are as given in Fig.~\ref{fig:pd_per}.}
\label{fig:pd_dur}
\end{figure}

Fig.~\ref{fig:pd_dur} shows a positive correlation between the absolute value of the period drift of the QPPs and the duration of the flare, with a Pearson correlation coefficient of 0.82. This relationship can likely be attributed to the fact that longer duration flares allow more time for any non-stationary QPP periods to evolve which leads to greater magnitude period drifts, in addition to the artificial correlation between average period and absolute period drift, seen in Fig.~\ref{fig:pd_per}. There is no noticeable difference between the relationship of flare duration to period drift magnitude for positive or negative period drifts. 

The period drift of all QPPs in the 98 flares may be visualised in Fig.~\ref{fig:per_arrow} (or explored in Table~\ref{tab:all_results} found in the Appendix). The periods of the QPPs are given in the horizontal axis, with bullet points indicating the period in the impulsive phase, and arrow heads indicating the period at the decay phase. Therefore arrows pointing right and coloured red indicate a positive period drift. Conversely blue arrows, pointing left, indicate a negative period drift. The period drift from a given flare is plotted against the corresponding flare's duration. The inset axes shows an enlarged region of the plot for flares with durations less than 2500~s. Flares with longer durations naturally allow for more time to evolve, leading to larger magnitude period drifts as discussed previously. The majority of results are clustered for flare durations less than 2500~s {($\sim$~40 minutes)}, with impulsive and decay phase periods of 40~s or less.

\begin{figure}
\centering
\includegraphics[width=\linewidth]{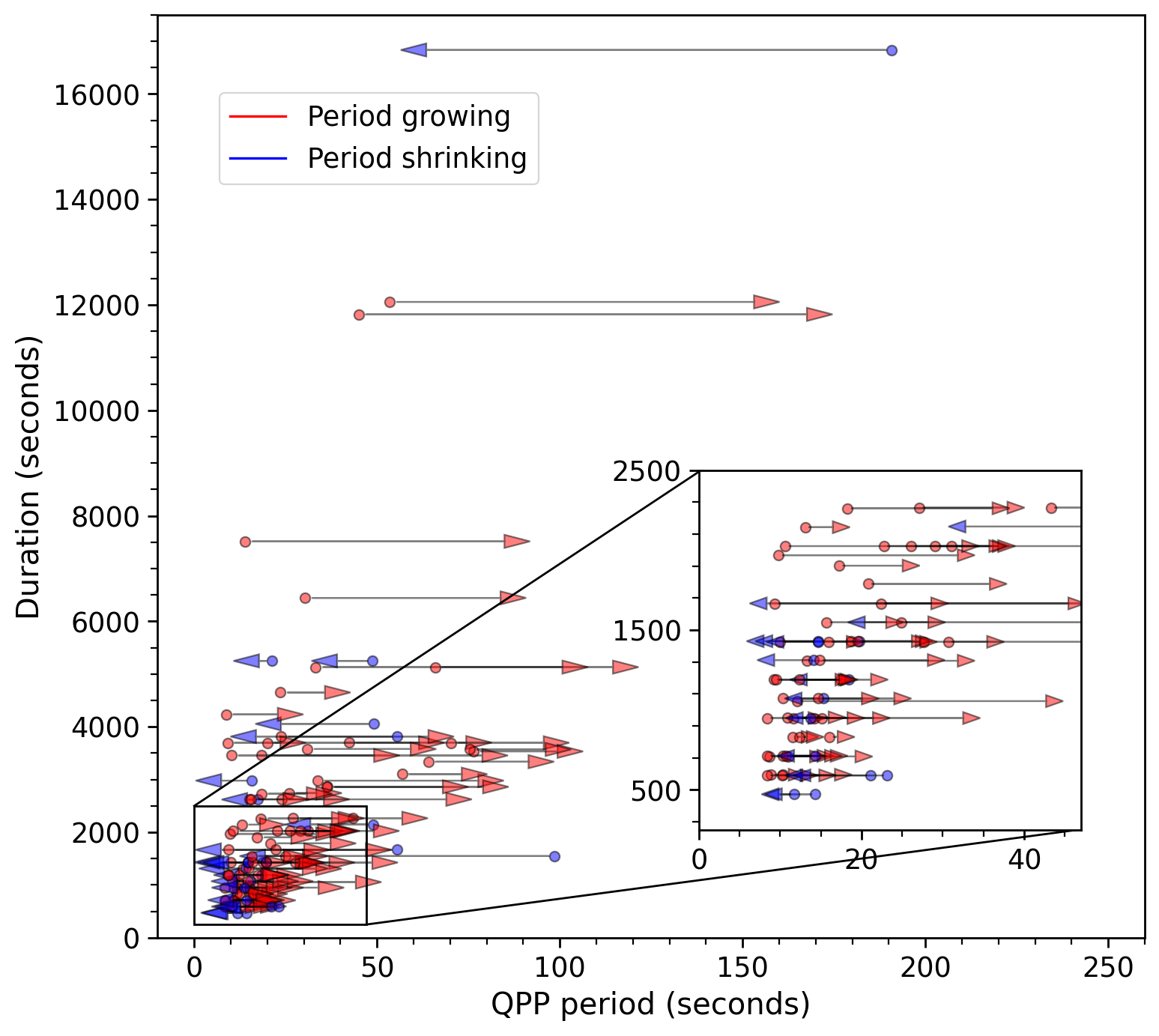}
\caption{Arrows show evolution of statistically significant periods in the impulsive and decay phases of 98 flares, with the arrow pointing from impulsive phase (indicated with a bullet point) to decay phase (arrow head). A period growth, i.e. a positive period drift, is shown in red and a negative period drift is given in blue. The period drift in the QPP is plotted against the flare's duration, both given in seconds.}
\label{fig:per_arrow}
\end{figure}

\begin{figure} b
\begin{subfigure}[b]{0.48\textwidth}
\centering
\includegraphics[width=.97\linewidth]{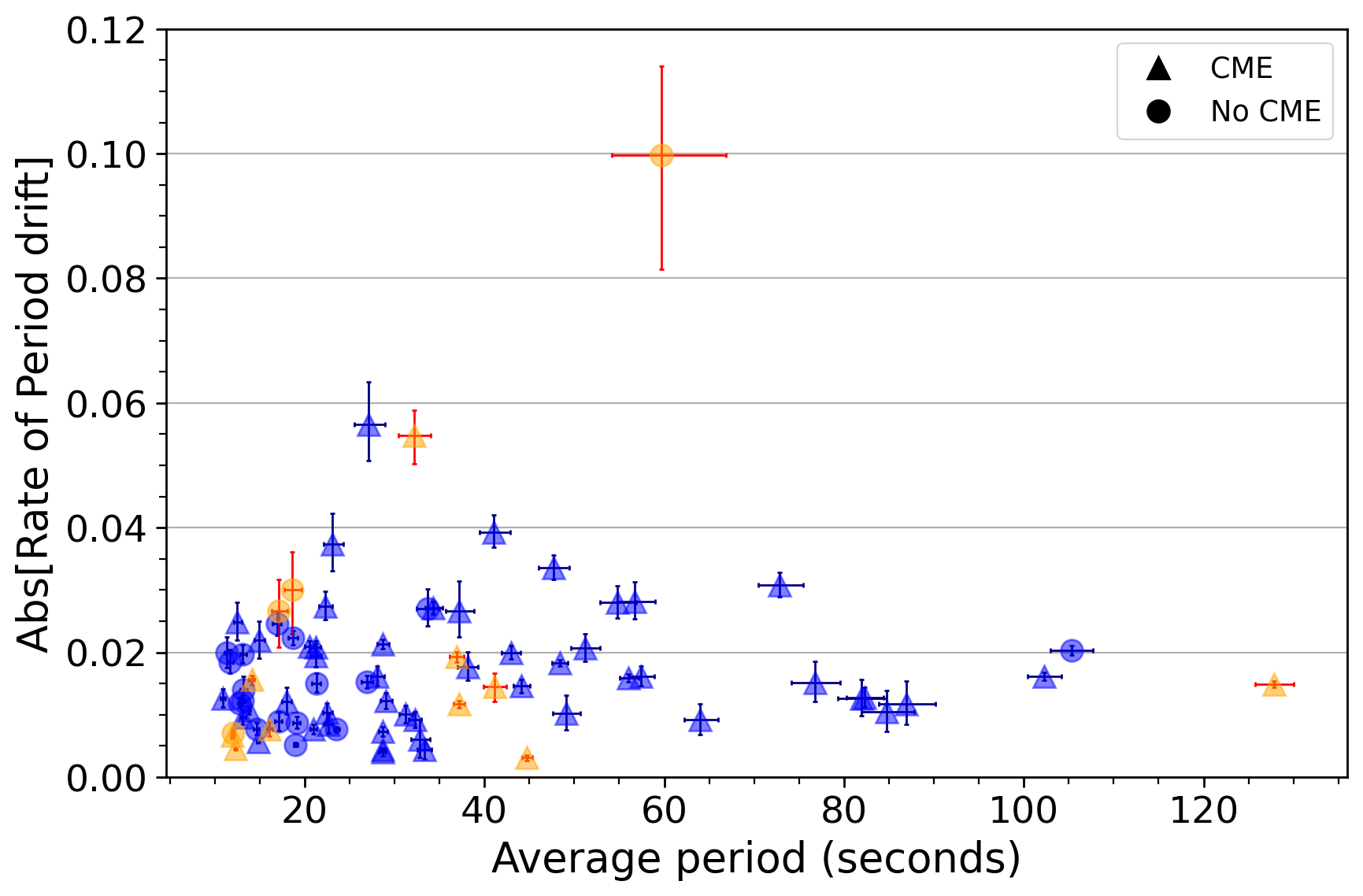}
\caption{}
\label{fig:pd_rate_scatter}
\end{subfigure}
\begin{subfigure}[b]{0.48\textwidth}
\centering
\includegraphics[width=.95\linewidth]{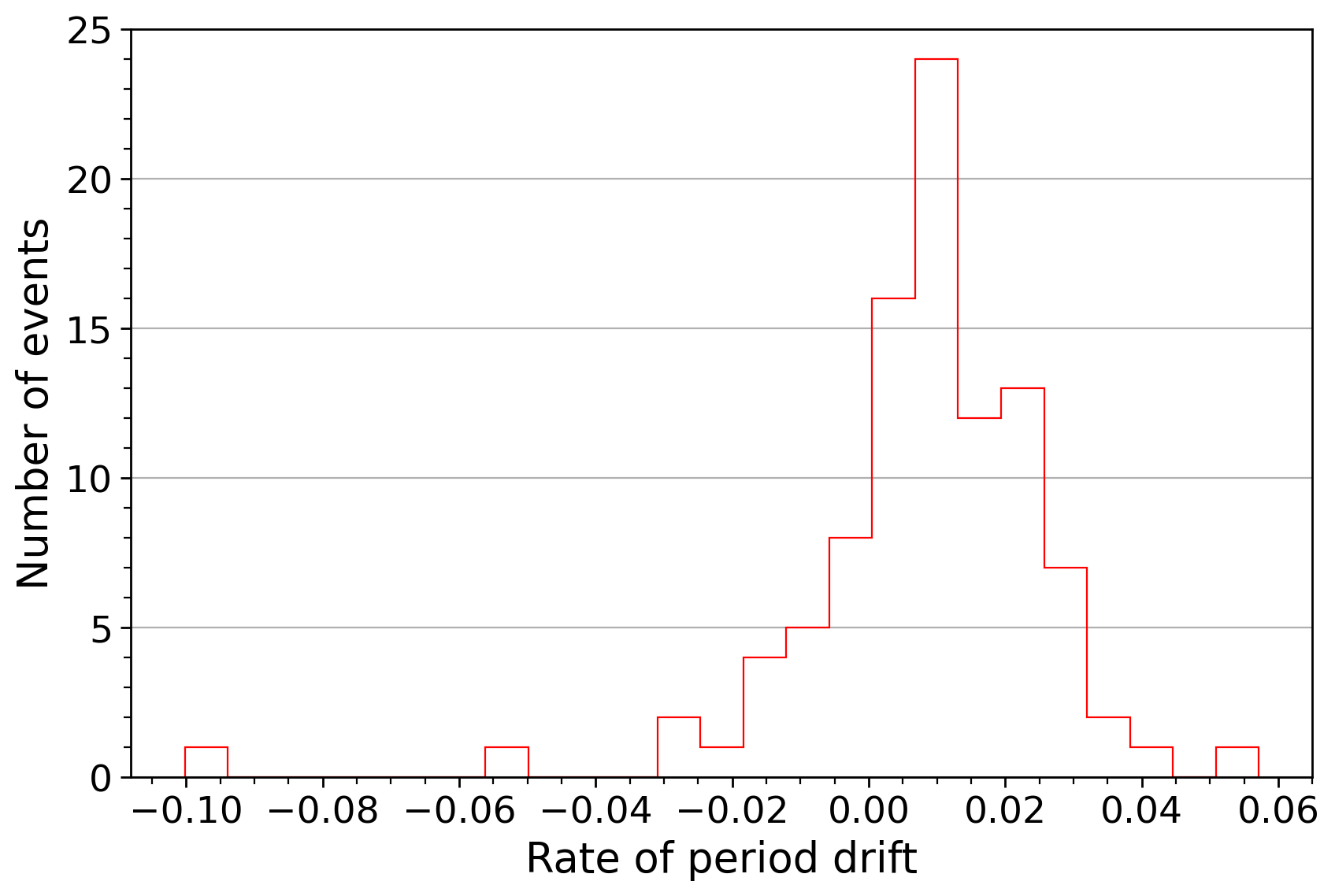}
\caption{}
\label{fig:pd_rate_hist}
\end{subfigure}
\caption{\textit{Top}: Scatter plot of the absolute magnitude of the rate of period drift, plotted against the average period of the QPP. \textit{Lower}: Histogram of rate of period drift.  The meanings of colours and symbols in the Top panel are as given in Fig.~\ref{fig:pd_per}.\label{fig:pd_rate}}
\end{figure}

We control for the duration of the flares and now examine the rate at which the QPP periods evolve. The rate of period drift is defined as the period drift divided by half the duration of the flare, and is therefore a unitless quantity. Fig.~\ref{fig:pd_rate} shows the distribution of the magnitude of the rate of period drift against average QPP period as a scatter plot (Top) and histogram (Lower). As can be seen, the rates of period drift have considerable scatter, although the absolute rate of period drift appears to cluster around $\sim$0.01 for average periods greater than 40~s, an effect that cannot be attributed to the selection criteria. Due to the selection criteria discussed in the methods section, the maximal possible absolute rate of period drift for the data used in this study is 1.4. The maximal rate of positive period drift seen in these results is 0.06 and the maximal rate of negative period drift is -0.1, although the majority of the rates of period drift are between0.02 and 0.03. There is no apparent correlation between the presence of a CME and the rate at which the QPP in the associated flare evolves. We also find that there is no correlation between the rate of period change and the flare energy, which suggests that QPP periods evolve at a rate independent of the peak flare energy. We also see the rate of period change to be uncorrelated with flare duration. This can be seen in Figs.~\ref{Appendix:energy_rate},~\ref{Appendix:duration_rate} in the Appendix.

\section{Discussion}\label{Discussion}

Firstly, we remind the reader of the biases and limitations of our study. All of the flaring events we examined had evidence of QPPs in the first place, detected by Fourier analysis. This biases the dataset towards QPPs that were stationary or slowly-evolving in periodicity, meaning that the results in this paper are likely to underestimate the population of QPPs undergoing rapid period evolution. We have chosen to split the flare into two phases, a choice which is ultimately arbitrary and done for convenience. This again biases the data and forces QPPs to be represented as stationary within an individual phase. It also neglects the possibility of QPPs which exist in e.g. only the impulsive or decay phase, or a shorter duration, which may be driven by entirely different generation mechanisms to the QPPs examined here. A more comprehensive study should look at QPP period evolution as a continuous process. It may be that any apparent period evolution is non-linear and follows some different schema. By repeating this analysis with some method that has time resolution, such as a continuous wavelet-transform (CWT) or Empirical Mode Decomposition (EMD) we may be able to uncover valuable information about the time evolution of the apparent period drifts. This may also be useful in discerning the generation mechanism(s) that are active in the appearance of these QPPs. 

As discussed earlier, a reader may be misled by these results into thinking that a single process is occurring in which the period is growing or shrinking. Instead it is possible that several periodicities exist at once, each generated by a separate QPP mechanism. A limitation of this work is that we only extract the period associated with the dominant peak from the FFT spectrum, ignoring additional potentially statistically significant peaks. In this paper, we associate the dominant periods in the FFT spectrum of each phase to produce a period drift, however this may not always be the most appropriate way to examine the change in instantaneous period of a QPP. For example it is possible for a given stationary periodicity to be present throughout the duration of the flare, and appear as the dominant peak in the FFT spectrum of the impulsive phase but as a secondary peak in the FFT spectrum of the decay phase due to an emergence of a secondary periodic process with greater amplitude. This may produce the appearance of a large magnitude period drift when both processes may in fact be stationary. However for the majority of the events assessed here (77/98, 79$\%$), both the FFT spectra of the impulsive and decay phases either resulted in dominant periods that were similar in magnitude (suggesting the direct evolution of a singular process) or produced only one peak in each phase that fulfilled the criteria discussed in Section~\ref{subsec:method} and appeared above the 95$\%$ confidence level. Therefore for these results the risks of drawing incorrect conclusions due to erroneously associated periodicities is low.

We have shown that the majority (81$\%$) of flaring events which have evidence of QPPs in both the impulsive and decay phases exhibit non-stationary behaviour. Although this sample is not strictly representative of the behaviour of QPPs en masse, due to the aforementioned biases in the data, the results discussed here are a strong indicator that we must consider non-stationarity to be a common property of QPPs and account for it in our methodology. If we search for QPPs by utilising methods that assume a stationary output, such as the FFT, we risk false-negative results where the non-stationarity of QPPs may cause spectral leakage. We also risk poorly categorising the behaviour of QPPs by assigning a single value for QPP period. This is important because different QPP mechanisms allow for the presence of non-stationarity in different ways and we must not omit the valuable data by treating the QPP periods as a fixed value if we are to determine what causes QPPs.

We also note the disparity in the proportion of flaring events showing a positive period drift (66$\%$) compared to those showing a negative period drift (14$\%$). This suggests an apparent growth in QPP period is more common than an apparent shrinkage, as previously reported in single event studies \citep[e.g.][]{hayes2016quasi, dennis2017detection, 2019Hayes}, and for a smaller statistical study \citep{simoes2015soft, Hayes_2020}.  We also note that most of the period drift that we observe is of small magnitude- most commonly between $\pm$10~s. 

The rates at which the QPPs evolved in period exist over the same ranges and in roughly the same populations for both growing and shrinking QPP periods, without any dependence on QPP average period or maximum flare energy.  We note that the presence of CMEs or peak flare energy seem to have no effect on whether the QPP periods grow or shrink, or the magnitudes of the period drifts. We see that longer duration flares are correlated with greater magnitude period drift. It is possible that other properties, such as CME speed or the magnetic configuration of the Active Region could play a role in determining if and how the QPP periods evolve. 

\section{Conclusions}\label{conclusion}

There is clear evidence that non-stationarity is a common phenomenon in QPPs observed in M- and X- class solar flares, with period growth appearing more common than period shrinkage. We must consider this when investigating flaring events for QPPs and be wary about how we assign values to QPP periodicities. It appears that most QPP that show non-stationarity evolve in period at similar rates. It is unlikely that the presence alone of CMEs, or the peak flare energy impacts the presence or magnitude of QPP period evolution. As seen in Table 1 of \citet[][]{2021Zimovets} there are many generation mechanisms (from all of the previously mentioned groupings) that have the potential to produce QPPs with non-stationary properties. In building a catalogue of QPPs that exhibit non-stationarity (see Table B1 in the appendix) future work may determine commonalities, such as the magnetic configuration of the flare site, which could be used to narrow down which mechanisms are responsible for driving non-stationary behaviour. Further work with spatial resolution of the flare site may be valuable in investigating the cause of QPP period evolution.

\section*{Acknowledgements}
L.A.H is supported by an ESA Research Fellowship. A-MB acknowledges support from the Science and Technology Facilities Council (STFC) consolidated grant ST/T000252/1. The CME catalogue is generated and maintained at the CDAW Data Center by NASA and The Catholic University of America in cooperation with the Naval Research Laboratory. SOHO is a project of international cooperation between ESA and NASA. This research was supported by the International Space Science Institute (ISSI) in Bern, through ISSI International Team project 527: Bridging New X-ray Observations and Advanced Models of Flare Variability: A Key to Understanding the Fundamentals of Flare Energy Release. This research made use of \textsc{sunpy} \citep{sunpy}, \textsc{matplotlib} \citep{matplotlib}, \textsc{numpy} \citep{numpy}, \textsc{pandas} \citep{pandas2}, and \textsc{scipy} \citep{scipy}.

\section*{Data availability}
The data used here are all publicly available. The GOES-XRS data is available online from NOAA (\url{ngdc.noaa.gov/stp/satellite/goes/index.html}) and the SOHO/LASCO CME catalogue can be found here (\url{cdaw.gsfc.nasa.gov/CME_list/}). The procedures described in Section~\ref{subsec:method} can be found in the following repository: \url{github.com/chloepugh/QPP-confidence-levels}. 




\bibliographystyle{mnras}
\bibliography{qppbib} 



\section{Appendix A}

We include Figs.~\ref{Appendix:energy_rate}, \ref{Appendix:duration_rate}, which show the absolute values of the rate of period drift against peak flare energy and flare duration respectively. No correlation is observed in either figure. This is expected for Fig.~\ref{Appendix:duration_rate} as we obtain the rate of period drift by dividing the period drift by the flare duration. Therefore we remove the duration dependence seen in Fig.~\ref{fig:pd_dur}.

\renewcommand{\thefigure}{A\arabic{figure}}
\setcounter{figure}{0}

\begin{figure}
\centering
\includegraphics[width=\linewidth]{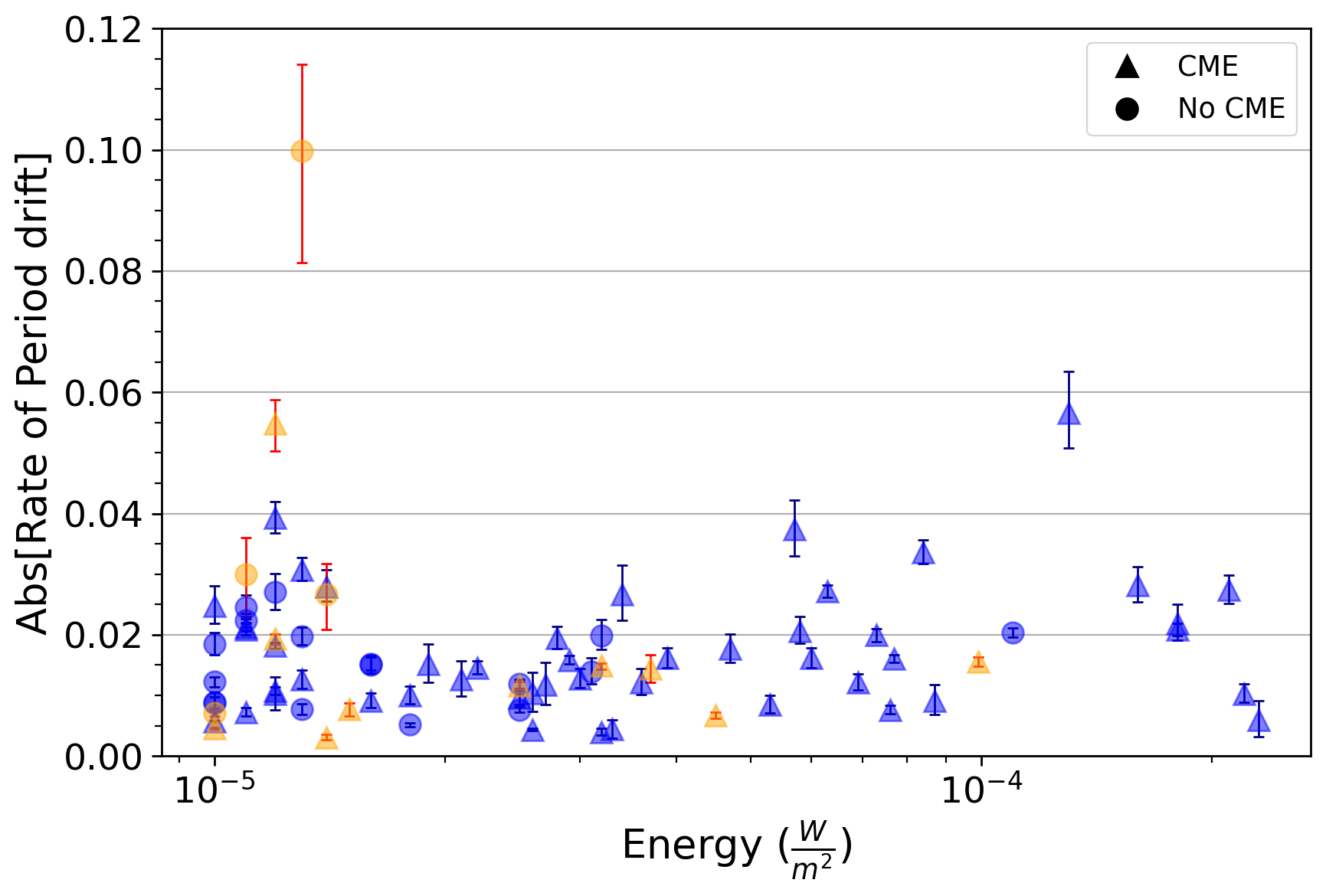}
\caption{Rate of period drift plotted against peak flare energy. The meanings of colours and symbols are as given in Fig.~\ref{fig:pd_per}.}
\label{Appendix:energy_rate}
\end{figure}

\begin{figure}
\centering
\includegraphics[width=\linewidth]{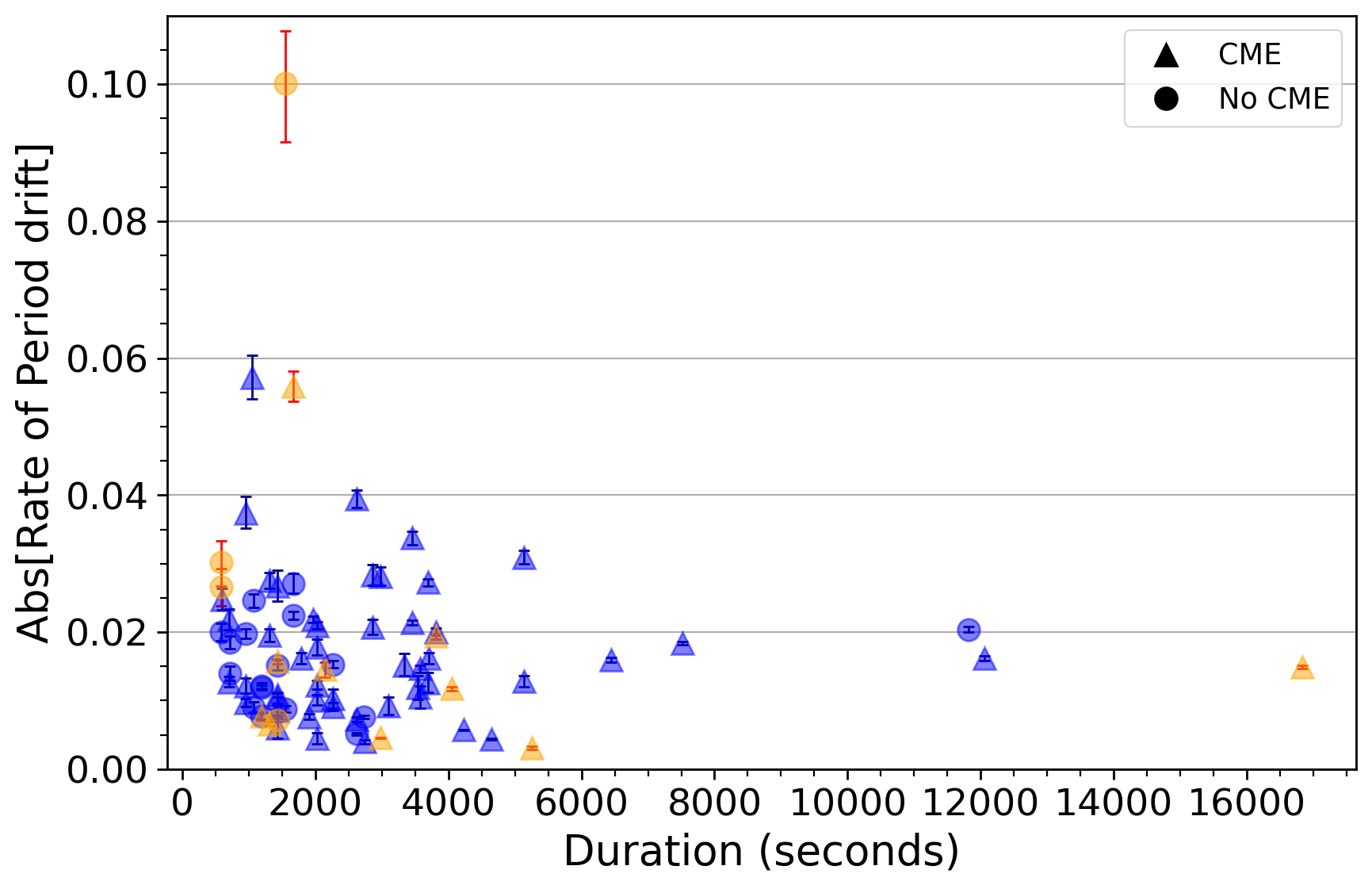}
\caption{Rate of period drift plotted against flare durations. The meanings of colours and symbols are as given in Fig.~\ref{fig:pd_per}.}
\label{Appendix:duration_rate}
\end{figure}

\section{Appendix B}
\renewcommand{\thetable}{B\arabic{table}}
\setcounter{table}{0}

Table~\ref{tab:all_results} shows a table of the flares examined in this study that fulfilled the three criteria outlined in Section~\ref{subsec:method} and the associated periods detected in the impulsive and decay phases. We also include the flare's GOES class, duration, and period drift. Table~\ref{tab:all_results} can be downloaded in comma-separated values format (csv) by accessing the following repository: \url{github.com/TaraAthem/Non-stationarity-in-QPPs}.

\onecolumn
\begin{table}
\caption{Num gives the unique ID number of each flaring event. Date refers to the date that the flaring event began and T$_{start}$, T$_{end}$, and T$_{peak}$ refer to the times used in this study corresponding to the start, end and peak flux of the flaring event, as measured in the long channel of \textit{GOES-15}. GOES class gives the flaring class of the event. The CME column is ticked with a check mark if the flaring event is associated with a Coronal Mass Ejection. Period$_{Avg}$ is the mean period of the flaring event, found by taking the average of the impulsive and decay phase periods (given in columns Period$_{Impulsive}$, Period$_{Decay}$). Period drift is given in the final column of the table.  \label{tab:all_results}} 
\label{tab:landscape}
\begin{tabular}{llllllccllll}
\hline\hline
\textbf{Num} &
  \textbf{Date} &
  \textbf{T$_{start}$} &
  \textbf{T$_{end}$} &
  \textbf{T$_{peak}$} &
  \textbf{Duration} &
  \textbf{GOES} &
  \textbf{CME} &
  \textbf{Period$_{Avg}$} &
  \textbf{Period$_{Impulsive}$} &
  \textbf{Period$_{Decay}$} &
  \textbf{Period drift} \\

  \textbf{} &
  \textbf{} &
  \textbf{} &
  \textbf{} &
  \textbf{} &
  (seconds) &
  \textbf{class} &
  \textbf{} &
  (seconds) &
  (seconds) &
  (seconds) &
  (seconds) \\
  \hline
01 &
  2011-02-14 &
  17:20:00 &
  17:31:55 &
  17:26:08 &
  715 &
  M2.2 &
  $\checkmark$ &
  12.2$\genfrac{}{}{0pt}{2}{+0.3}{-0.3}$ &
  10.3$\genfrac{}{}{0pt}{2}{+0.3}{-0.3}$ &
  14.1$\genfrac{}{}{0pt}{2}{+0.6}{-0.5}$ &
  3.7$\genfrac{}{}{0pt}{2}{+0.7}{-0.6}$ $\dagger$ \\
02 &
  2011-02-15 &
  01:44:00 &
  02:07:56 &
  01:56:44 &
  1436 &
  X2.2 &
  $\checkmark$ &
  22.5$\genfrac{}{}{0pt}{2}{+0.6}{-0.5}$ &
  18.8$\genfrac{}{}{0pt}{2}{+0.5}{-0.5}$ &
  26.2$\genfrac{}{}{0pt}{2}{+1.0}{-0.9}$ &
  7.4$\genfrac{}{}{0pt}{2}{+1.1}{-1.0}$ \\
03 &
  2011-02-18 &
  20:56:00 &
  21:11:53 &
  21:03:58 &
  953 &
  M1.3 &
   &
  13.1$\genfrac{}{}{0pt}{2}{+0.4}{-0.3}$ &
  8.4$\genfrac{}{}{0pt}{2}{+0.2}{-0.1}$ &
  17.8$\genfrac{}{}{0pt}{2}{+0.7}{-0.6}$ &
  9.4$\genfrac{}{}{0pt}{2}{+0.7}{-0.7}$ \\
04 &
  2011-04-22 &
  04:35:00 &
  05:18:56 &
  04:56:43 &
  2636 &
  M1.8 &
   &
  19$\genfrac{}{}{0pt}{2}{+0.2}{-0.2}$ &
  15.6$\genfrac{}{}{0pt}{2}{+0.2}{-0.2}$ &
  22.4$\genfrac{}{}{0pt}{2}{+0.4}{-0.4}$ &
  6.8$\genfrac{}{}{0pt}{2}{+0.4}{-0.4}$ \\
05 &
  2011-05-29 &
  10:08:00 &
  10:57:54 &
  10:33:15 &
  2994 &
  M1.4 &
  $\checkmark$ &
  54.7$\genfrac{}{}{0pt}{2}{+2.1}{-1.9}$ &
  33.8$\genfrac{}{}{0pt}{2}{+0.8}{-0.7}$ &
  75.7$\genfrac{}{}{0pt}{2}{+4.0}{-3.6}$ &
  41.9$\genfrac{}{}{0pt}{2}{+4.1}{-3.7}$ \\
06 &
  2011-08-03 &
  13:17:00 &
  14:18:57 &
  13:47:56 &
  3717 &
  M6.0 &
  $\checkmark$ &
  57.4$\genfrac{}{}{0pt}{2}{+1.5}{-1.4}$ &
  42.4$\genfrac{}{}{0pt}{2}{+1.0}{-0.9}$ &
  72.4$\genfrac{}{}{0pt}{2}{+2.9}{-2.7}$ &
  30$\genfrac{}{}{0pt}{2}{+3.1}{-2.9}$ \\
07 &
  2011-09-07 &
  22:32:00 &
  22:43:52 &
  22:38:44 &
  712 &
  X1.8 &
  $\checkmark$ &
  14.9$\genfrac{}{}{0pt}{2}{+0.6}{-0.5}$ &
  11$\genfrac{}{}{0pt}{2}{+0.4}{-0.3}$ &
  18.8$\genfrac{}{}{0pt}{2}{+1.0}{-0.9}$ &
  7.8$\genfrac{}{}{0pt}{2}{+1.1}{-1.0}$ \\
08 &
  2011-09-10 &
  07:18:00 &
  08:01:56 &
  07:40:43 &
  2636 &
  M1.1 &
  $\checkmark$ &
  28.7$\genfrac{}{}{0pt}{2}{+0.5}{-0.5}$ &
  23.9$\genfrac{}{}{0pt}{2}{+0.4}{-0.4}$ &
  33.5$\genfrac{}{}{0pt}{2}{+0.9}{-0.8}$ &
  9.6$\genfrac{}{}{0pt}{2}{+1.0}{-0.9}$ \\
09 &
  2011-09-23 &
  23:48:00 &
  00:03:54 * &
  23:55:41 &
  954 &
  M1.9 &
  $\checkmark$ &
  12.3$\genfrac{}{}{0pt}{2}{+0.2}{-0.2}$ &
  11.6$\genfrac{}{}{0pt}{2}{+0.3}{-0.3}$ &
  13$\genfrac{}{}{0pt}{2}{+0.4}{-0.3}$ &
  1.4$\genfrac{}{}{0pt}{2}{+0.5}{-0.4}$ $\dagger$ \\
10 &
  2011-09-24 &
  17:19:00 &
  17:30:55 &
  17:25:33 &
  715 &
  M3.1 &
   &
  13.2$\genfrac{}{}{0pt}{2}{+0.4}{-0.4}$ &
  10.7$\genfrac{}{}{0pt}{2}{+0.3}{-0.3}$ &
  15.7$\genfrac{}{}{0pt}{2}{+0.7}{-0.7}$ &
  5$\genfrac{}{}{0pt}{2}{+0.8}{-0.7}$ \\
11 &
  2011-09-25 &
  09:25:00 &
  09:44:54 &
  09:35:56 &
  1194 &
  M1.5 &
  $\checkmark$ &
  16.1$\genfrac{}{}{0pt}{2}{+0.3}{-0.3}$ &
  18.4$\genfrac{}{}{0pt}{2}{+0.6}{-0.6}$ &
  13.8$\genfrac{}{}{0pt}{2}{+0.3}{-0.3}$ &
  -4.6$\genfrac{}{}{0pt}{2}{+0.7}{-0.6}$ \\
12 &
  2011-10-01 &
  08:56:00 &
  11:01:52 &
  09:59:21 &
  7552 &
  M1.2 &
  $\checkmark$ &
  48.4$\genfrac{}{}{0pt}{2}{+0.9}{-0.9}$ &
  13.9$\genfrac{}{}{0pt}{2}{+0.1}{-0.1}$ &
  82.9$\genfrac{}{}{0pt}{2}{+1.9}{-1.8}$ &
  68.9$\genfrac{}{}{0pt}{2}{+1.9}{-1.8}$ \\
13 &
  2011-11-05 &
  20:31:00 &
  20:44:53 &
  20:38:34 &
  833 &
  M1.8 &
   &
  12.6$\genfrac{}{}{0pt}{2}{+0.3}{-0.3}$ &
  12.4$\genfrac{}{}{0pt}{2}{+0.4}{-0.4}$ &
  12.8$\genfrac{}{}{0pt}{2}{+0.4}{-0.4}$ &
  0.4$\genfrac{}{}{0pt}{2}{+0.6}{-0.5}$ $\dagger$ \\
14 &
  2012-01-17 &
  04:41:00 &
  05:04:56 &
  04:53:40 &
  1436 &
  M1.0 &
   &
  12.1$\genfrac{}{}{0pt}{2}{+0.2}{-0.2}$ &
  14.6$\genfrac{}{}{0pt}{2}{+0.3}{-0.3}$ &
  9.5$\genfrac{}{}{0pt}{2}{+0.1}{-0.1}$ &
  -5.1$\genfrac{}{}{0pt}{2}{+0.3}{-0.3}$ \\
15 &
  2012-01-19 &
  13:44:00 &
  18:25:54 &
  16:03:17 &
  16914 &
  M3.2 &
  $\checkmark$ &
  127.9$\genfrac{}{}{0pt}{2}{+2.2}{-2.1}$ &
  190.7$\genfrac{}{}{0pt}{2}{+4.4}{-4.2}$ &
  65$\genfrac{}{}{0pt}{2}{+0.5}{-0.5}$ &
  -125.8$\genfrac{}{}{0pt}{2}{+4.4}{-4.2}$ \\
16 &
  2012-03-02 &
  17:29:00 &
  18:02:55 &
  17:46:26 &
  2035 &
  M3.3 &
  $\checkmark$ &
  33.4$\genfrac{}{}{0pt}{2}{+0.8}{-0.8}$ &
  31.1$\genfrac{}{}{0pt}{2}{+1.0}{-0.9}$ &
  35.6$\genfrac{}{}{0pt}{2}{+1.3}{-1.2}$ &
  4.5$\genfrac{}{}{0pt}{2}{+1.6}{-1.5}$ \\
17 &
  2012-03-05 &
  02:30:00 &
  05:47:56 &
  04:08:34 &
  11876 &
  X1.1 &
   &
  105.3$\genfrac{}{}{0pt}{2}{+2.4}{-2.3}$ &
  45$\genfrac{}{}{0pt}{2}{+0.3}{-0.3}$ &
  165.6$\genfrac{}{}{0pt}{2}{+4.8}{-4.5}$ &
  120.6$\genfrac{}{}{0pt}{2}{+4.8}{-4.5}$ \\
18 &
  2012-03-07 &
  01:05:00 &
  01:22:47 &
  01:15:25 &
  1067 &
  X1.3 &
  $\checkmark$ &
  27.1$\genfrac{}{}{0pt}{2}{+1.8}{-1.6}$ &
  12$\genfrac{}{}{0pt}{2}{+0.3}{-0.3}$ &
  42.2$\genfrac{}{}{0pt}{2}{+3.6}{-3.1}$ &
  30.2$\genfrac{}{}{0pt}{2}{+3.6}{-3.1}$ \\
19 &
  2012-03-09 &
  03:22:00 &
  04:23:53 &
  03:53:19 &
  3713 &
  M6.3 &
  $\checkmark$ &
  34.3$\genfrac{}{}{0pt}{2}{+1.0}{-0.9}$ &
  9.1$\genfrac{}{}{0pt}{2}{+0.1}{-0.1}$ &
  59.5$\genfrac{}{}{0pt}{2}{+2.0}{-1.8}$ &
  50.4$\genfrac{}{}{0pt}{2}{+2.0}{-1.8}$ \\
20 &
  2012-03-10 &
  17:15:00 &
  18:12:52 &
  17:43:55 &
  3472 &
  M8.4 &
  $\checkmark$ &
  47.7$\genfrac{}{}{0pt}{2}{+1.8}{-1.6}$ &
  18.5$\genfrac{}{}{0pt}{2}{+0.2}{-0.2}$ &
  76.8$\genfrac{}{}{0pt}{2}{+3.6}{-3.3}$ &
  58.3$\genfrac{}{}{0pt}{2}{+3.6}{-3.3}$ \\
21 &
  2012-05-06 &
  01:12:00 &
  01:23:55 &
  01:18:05 &
  715 &
  M1.1 &
   &
  13.3$\genfrac{}{}{0pt}{2}{+0.4}{-0.3}$ &
  14.3$\genfrac{}{}{0pt}{2}{+0.6}{-0.6}$ &
  12.2$\genfrac{}{}{0pt}{2}{+0.4}{-0.4}$ &
  -2.1$\genfrac{}{}{0pt}{2}{+0.7}{-0.7}$ $\dagger$ \\
22 &
  2012-05-07 &
  14:03:00 &
  14:58:53 &
  14:31:18 &
  3353 &
  M1.9 &
  $\checkmark$ &
  76.8$\genfrac{}{}{0pt}{2}{+2.8}{-2.6}$ &
  64.1$\genfrac{}{}{0pt}{2}{+2.5}{-2.4}$ &
  89.4$\genfrac{}{}{0pt}{2}{+5.0}{-4.5}$ &
  25.4$\genfrac{}{}{0pt}{2}{+5.6}{-5.1}$ \\
23 &
  2012-05-09 &
  21:01:00 &
  21:08:53 &
  21:05:22 &
  473 &
  M4.1 &
  $\checkmark$ &
  12.3$\genfrac{}{}{0pt}{2}{+0.5}{-0.5}$ &
  14.3$\genfrac{}{}{0pt}{2}{+0.9}{-0.8}$ &
  10.3$\genfrac{}{}{0pt}{2}{+0.5}{-0.4}$ &
  -4$\genfrac{}{}{0pt}{2}{+1.0}{-0.9}$ $\dagger$ \\
24 &
  2012-05-10 &
  04:11:00 &
  04:24:54 &
  04:17:50 &
  834 &
  M5.7 &
  $\checkmark$ &
  16.3$\genfrac{}{}{0pt}{2}{+0.5}{-0.4}$ &
  16.1$\genfrac{}{}{0pt}{2}{+0.6}{-0.6}$ &
  16.6$\genfrac{}{}{0pt}{2}{+0.7}{-0.6}$ &
  0.5$\genfrac{}{}{0pt}{2}{+0.9}{-0.9}$ \\
25 &
  2012-07-19 &
  04:17:00 &
  07:38:55 &
  05:57:51 &
  12115 &
  M7.7 &
  $\checkmark$ &
  102.3$\genfrac{}{}{0pt}{2}{+1.9}{-1.9}$ &
  53.5$\genfrac{}{}{0pt}{2}{+0.5}{-0.5}$ &
  151.1$\genfrac{}{}{0pt}{2}{+3.9}{-3.7}$ &
  97.6$\genfrac{}{}{0pt}{2}{+3.9}{-3.7}$ \\
26 &
  2012-07-30 &
  15:39:00 &
  15:56:55 &
  15:48:28 &
  1075 &
  M1.1 &
   &
  16.9$\genfrac{}{}{0pt}{2}{+0.5}{-0.5}$ &
  10.3$\genfrac{}{}{0pt}{2}{+0.2}{-0.2}$ &
  23.5$\genfrac{}{}{0pt}{2}{+1.1}{-1.0}$ &
  13.2$\genfrac{}{}{0pt}{2}{+1.1}{-1.0}$ \\
27 &
  2012-08-11 &
  11:55:00 &
  12:44:54 &
  12:19:52 &
  2994 &
  M1.0 &
  $\checkmark$ &
  12.3$\genfrac{}{}{0pt}{2}{+0.1}{-0.1}$ &
  15.7$\genfrac{}{}{0pt}{2}{+0.2}{-0.2}$ &
  8.9$\genfrac{}{}{0pt}{2}{+0.1}{-0.1}$ &
  -6.8$\genfrac{}{}{0pt}{2}{+0.2}{-0.2}$ \\
28 &
  2012-08-30 &
  12:02:00 &
  12:19:55 &
  12:11:36 &
  1075 &
  M1.3 &
   &
  14.2$\genfrac{}{}{0pt}{2}{+0.3}{-0.3}$ &
  15.3$\genfrac{}{}{0pt}{2}{+0.4}{-0.4}$ &
  13.1$\genfrac{}{}{0pt}{2}{+0.3}{-0.3}$ &
  -2.2$\genfrac{}{}{0pt}{2}{+0.6}{-0.5}$ $\dagger$ \\
29 &
  2012-09-30 &
  04:27:00 &
  04:38:52 &
  04:33:01 &
  712 &
  M1.3 &
  $\checkmark$ &
  10.9$\genfrac{}{}{0pt}{2}{+0.3}{-0.3}$ &
  8.6$\genfrac{}{}{0pt}{2}{+0.2}{-0.2}$ &
  13.2$\genfrac{}{}{0pt}{2}{+0.5}{-0.5}$ &
  4.5$\genfrac{}{}{0pt}{2}{+0.6}{-0.5}$ \\
30 &
  2012-10-08 &
  11:05:00 &
  11:28:53 &
  11:16:56 &
  1433 &
  M2.3 &
   &
  12.8$\genfrac{}{}{0pt}{2}{+0.2}{-0.2}$ &
  14.6$\genfrac{}{}{0pt}{2}{+0.3}{-0.3}$ &
  10.9$\genfrac{}{}{0pt}{2}{+0.2}{-0.2}$ &
  -3.7$\genfrac{}{}{0pt}{2}{+0.3}{-0.3}$ $\dagger$ \\
31 &
  2012-11-27 &
  15:52:00 &
  16:01:53 &
  15:57:35 &
  593 &
  M1.6 &
   &
  11.1$\genfrac{}{}{0pt}{2}{+0.3}{-0.3}$ &
  10.3$\genfrac{}{}{0pt}{2}{+0.4}{-0.3}$ &
  12$\genfrac{}{}{0pt}{2}{+0.5}{-0.5}$ &
  1.8$\genfrac{}{}{0pt}{2}{+0.6}{-0.6}$ $\dagger$ \\
32 &
  2013-05-03 &
  17:24:00 &
  17:39:53 &
  17:32:13 &
  953 &
  M5.7 &
  $\checkmark$ &
  23.1$\genfrac{}{}{0pt}{2}{+1.2}{-1.0}$ &
  14.2$\genfrac{}{}{0pt}{2}{+0.4}{-0.4}$ &
  32$\genfrac{}{}{0pt}{2}{+2.3}{-2.0}$ &
  17.8$\genfrac{}{}{0pt}{2}{+2.3}{-2.1}$ \\
33 &
  2013-06-05 &
  08:14:00 &
  09:39:55 &
  08:57:28 &
  5155 &
  M1.3 &
  $\checkmark$ &
  72.9$\genfrac{}{}{0pt}{2}{+2.6}{-2.4}$ &
  33.2$\genfrac{}{}{0pt}{2}{+0.4}{-0.4}$ &
  112.5$\genfrac{}{}{0pt}{2}{+5.1}{-4.7}$ &
  79.3$\genfrac{}{}{0pt}{2}{+5.2}{-4.7}$ \\
34 &
  2013-08-17 &
  18:49:00 &
  20:16:55 &
  19:33:47 &
  5275 &
  M1.4 &
  $\checkmark$ &
  44.7$\genfrac{}{}{0pt}{2}{+0.6}{-0.5}$ &
  48.8$\genfrac{}{}{0pt}{2}{+0.9}{-0.9}$ &
  40.6$\genfrac{}{}{0pt}{2}{+0.6}{-0.6}$ &
  -8.2$\genfrac{}{}{0pt}{2}{+1.1}{-1.1}$ \\
35 &
  2013-10-13 &
  00:12:00 &
  01:13:54 &
  00:43:36 &
  3714 &
  M1.7 &
  $\checkmark$ &
  20.8$\genfrac{}{}{0pt}{2}{+0.2}{-0.2}$ &
  20$\genfrac{}{}{0pt}{2}{+0.2}{-0.2}$ &
  21.6$\genfrac{}{}{0pt}{2}{+0.3}{-0.2}$ &
  1.6$\genfrac{}{}{0pt}{2}{+0.3}{-0.3}$ $\dagger$ \\
36 &
  2013-10-17 &
  15:09:00 &
  16:12:53 &
  15:41:00 &
  3833 &
  M1.2 &
  $\checkmark$ &
  36.9$\genfrac{}{}{0pt}{2}{+0.8}{-0.8}$ &
  55.4$\genfrac{}{}{0pt}{2}{+1.6}{-1.6}$ &
  18.4$\genfrac{}{}{0pt}{2}{+0.2}{-0.2}$ &
  -37$\genfrac{}{}{0pt}{2}{+1.7}{-1.6}$ \\
37 &
  2013-10-28 &
  01:41:00 &
  02:24:53 &
  02:02:57 &
  2633 &
  X1.0 &
  $\checkmark$ &
  16.7$\genfrac{}{}{0pt}{2}{+0.2}{-0.1}$ &
  17.5$\genfrac{}{}{0pt}{2}{+0.2}{-0.2}$ &
  16$\genfrac{}{}{0pt}{2}{+0.2}{-0.2}$ &
  -1.5$\genfrac{}{}{0pt}{2}{+0.3}{-0.3}$ $\dagger$ \\
38 &
  2013-10-29 &
  21:42:00 &
  22:05:53 &
  21:54:30 &
  1433 &
  X2.3 &
  $\checkmark$ &
  32.8$\genfrac{}{}{0pt}{2}{+1.1}{-1.0}$ &
  30.7$\genfrac{}{}{0pt}{2}{+1.4}{-1.3}$ &
  34.9$\genfrac{}{}{0pt}{2}{+1.8}{-1.6}$ &
  4.3$\genfrac{}{}{0pt}{2}{+2.3}{-2.1}$ \\
39 &
  2013-11-05 &
  18:08:00 &
  18:17:56 &
  18:12:57 &
  596 &
  M1.0 &
  $\checkmark$ &
  12.5$\genfrac{}{}{0pt}{2}{+0.5}{-0.4}$ &
  8.8$\genfrac{}{}{0pt}{2}{+0.3}{-0.3}$ &
  16.2$\genfrac{}{}{0pt}{2}{+0.9}{-0.8}$ &
  7.4$\genfrac{}{}{0pt}{2}{+1.0}{-0.9}$ \\
40 &
  2013-11-21 &
  10:52:00 &
  11:29:55 &
  11:11:07 &
  2275 &
  M1.2 &
  $\checkmark$ &
  49.1$\genfrac{}{}{0pt}{2}{+1.6}{-1.5}$ &
  43.3$\genfrac{}{}{0pt}{2}{+1.7}{-1.6}$ &
  54.9$\genfrac{}{}{0pt}{2}{+2.8}{-2.5}$ &
  11.6$\genfrac{}{}{0pt}{2}{+3.3}{-3.0}$ \\
41 &
  2013-12-07 &
  07:17:00 &
  07:40:53 &
  07:29:41 &
  1433 &
  M1.2 &
  $\checkmark$ &
  13.8$\genfrac{}{}{0pt}{2}{+0.2}{-0.2}$ &
  9.9$\genfrac{}{}{0pt}{2}{+0.1}{-0.1}$ &
  17.7$\genfrac{}{}{0pt}{2}{+0.4}{-0.4}$ &
  7.7$\genfrac{}{}{0pt}{2}{+0.5}{-0.4}$ \\
42 &
  2013-12-31 &
  21:45:00 &
  22:10:56 &
  21:58:07 &
  1556 &
  M6.4 &
  $\checkmark$ &
  26.3$\genfrac{}{}{0pt}{2}{+0.7}{-0.6}$ &
  24.9$\genfrac{}{}{0pt}{2}{+0.8}{-0.8}$ &
  27.7$\genfrac{}{}{0pt}{2}{+1.0}{-1.0}$ &
  2.8$\genfrac{}{}{0pt}{2}{+1.3}{-1.2}$ $\dagger$ \\
43 &
  2014-01-01 &
  18:40:00 &
  19:04:10 &
  18:52:03 &
  1450 &
  M9.9 &
  $\checkmark$ &
  14.1$\genfrac{}{}{0pt}{2}{+0.3}{-0.3}$ &
  19.7$\genfrac{}{}{0pt}{2}{+0.6}{-0.5}$ &
  8.5$\genfrac{}{}{0pt}{2}{+0.1}{-0.1}$ &
  -11.3$\genfrac{}{}{0pt}{2}{+0.6}{-0.5}$ \\
44 &
  2014-01-08 &
  03:39:00 &
  03:54:54 &
  03:47:45 &
  954 &
  M3.6 &
  $\checkmark$ &
  18$\genfrac{}{}{0pt}{2}{+0.5}{-0.5}$ &
  15.1$\genfrac{}{}{0pt}{2}{+0.5}{-0.5}$ &
  20.9$\genfrac{}{}{0pt}{2}{+1.0}{-0.9}$ &
  5.8$\genfrac{}{}{0pt}{2}{+1.1}{-1.0}$ \\
45 &
  2014-01-30 &
  07:54:00 &
  08:28:17 &
  08:10:51 &
  2057 &
  M1.1 &
  $\checkmark$ &
  20.6$\genfrac{}{}{0pt}{2}{+0.5}{-0.5}$ &
  9.8$\genfrac{}{}{0pt}{2}{+0.1}{-0.1}$ &
  31.3$\genfrac{}{}{0pt}{2}{+1.0}{-0.9}$ &
  21.5$\genfrac{}{}{0pt}{2}{+1.0}{-0.9}$ \\
46 &
  2014-02-11 &
  16:34:00 &
  17:07:54 &
  16:51:43 &
  2034 &
  M1.8 &
  $\checkmark$ &
  31.2$\genfrac{}{}{0pt}{2}{+0.8}{-0.7}$ &
  26.1$\genfrac{}{}{0pt}{2}{+0.7}{-0.7}$ &
  36.3$\genfrac{}{}{0pt}{2}{+1.3}{-1.3}$ &
  10.2$\genfrac{}{}{0pt}{2}{+1.5}{-1.4}$ \\
47 &
  2014-02-24 &
  11:03:00 &
  11:31:20 &
  11:17:07 &
  1700 &
  M1.2 &
  $\checkmark$ &
  32.2$\genfrac{}{}{0pt}{2}{+1.9}{-1.7}$ &
  55.5$\genfrac{}{}{0pt}{2}{+3.9}{-3.4}$ &
  8.8$\genfrac{}{}{0pt}{2}{+0.1}{-0.1}$ &
  -46.6$\genfrac{}{}{0pt}{2}{+3.9}{-3.4}$ \\
48 &
  2014-03-10 &
  04:02:00 &
  04:13:55 &
  04:08:17 &
  715 &
  M1.0 &
   &
  11.7$\genfrac{}{}{0pt}{2}{+0.3}{-0.3}$ &
  8.4$\genfrac{}{}{0pt}{2}{+0.2}{-0.2}$ &
  15$\genfrac{}{}{0pt}{2}{+0.7}{-0.6}$ &
  6.6$\genfrac{}{}{0pt}{2}{+0.7}{-0.6}$ \\
49 &
  2014-03-12 &
  10:55:00 &
  11:14:54 &
  11:05:09 &
  1194 &
  M2.5 &
   &
  12.7$\genfrac{}{}{0pt}{2}{+0.2}{-0.2}$ &
  9.2$\genfrac{}{}{0pt}{2}{+0.1}{-0.1}$ &
  16.3$\genfrac{}{}{0pt}{2}{+0.5}{-0.4}$ &
  7.1$\genfrac{}{}{0pt}{2}{+0.5}{-0.5}$ \\
50 &
  2014-04-18 &
  12:31:00 &
  13:34:52 &
  13:02:58 &
  3832 &
  M7.3 &
  $\checkmark$ &
  42.9$\genfrac{}{}{0pt}{2}{+1.1}{-1.0}$ &
  23.8$\genfrac{}{}{0pt}{2}{+0.3}{-0.3}$ &
  62.1$\genfrac{}{}{0pt}{2}{+2.1}{-1.9}$ &
  38.2$\genfrac{}{}{0pt}{2}{+2.1}{-2.0}$ \\
51 &
  2014-05-07 &
  16:07:00 &
  16:50:56 &
  16:29:08 &
  2636 &
  M1.2 &
  $\checkmark$ &
  41.1$\genfrac{}{}{0pt}{2}{+1.8}{-1.6}$ &
  15.2$\genfrac{}{}{0pt}{2}{+0.2}{-0.2}$ &
  66.9$\genfrac{}{}{0pt}{2}{+3.6}{-3.2}$ &
  51.8$\genfrac{}{}{0pt}{2}{+3.6}{-3.2}$ \\
52 &
  2014-06-12 &
  04:14:00 &
  04:27:55 &
  04:21:17 &
  835 &
  M2.0 &
  $\checkmark$ &
  11.8$\genfrac{}{}{0pt}{2}{+0.2}{-0.2}$ &
  11.6$\genfrac{}{}{0pt}{2}{+0.3}{-0.3}$ &
  12.1$\genfrac{}{}{0pt}{2}{+0.4}{-0.3}$ &
  0.6$\genfrac{}{}{0pt}{2}{+0.5}{-0.5}$ $\dagger$ \\
53 &
  2014-06-12 &
  18:03:00 &
  18:22:54 &
  18:13:54 &
  1194 &
  M1.3 &
   &
  14.7$\genfrac{}{}{0pt}{2}{+0.3}{-0.3}$ &
  12.3$\genfrac{}{}{0pt}{2}{+0.3}{-0.2}$ &
  17$\genfrac{}{}{0pt}{2}{+0.5}{-0.5}$ &
  4.6$\genfrac{}{}{0pt}{2}{+0.6}{-0.5}$ \\
54 &
  2014-06-15 &
  11:10:00 &
  12:07:53 &
  11:39:34 &
  3473 &
  M1.1 &
  $\checkmark$ &
  28.7$\genfrac{}{}{0pt}{2}{+0.7}{-0.6}$ &
  10.2$\genfrac{}{}{0pt}{2}{+0.1}{-0.1}$ &
  47.2$\genfrac{}{}{0pt}{2}{+1.3}{-1.2}$ &
  37$\genfrac{}{}{0pt}{2}{+1.3}{-1.3}$ \\
55 &
  2014-07-10 &
  22:29:00 &
  22:38:56 &
  22:34:15 &
  596 &
  M1.5 &
  $\checkmark$ &
  10.4$\genfrac{}{}{0pt}{2}{+0.3}{-0.2}$ &
  10.3$\genfrac{}{}{0pt}{2}{+0.4}{-0.3}$ &
  10.5$\genfrac{}{}{0pt}{2}{+0.4}{-0.4}$ &
  0.2$\genfrac{}{}{0pt}{2}{+0.5}{-0.5}$ $\dagger$ \\
56 &
  2014-08-21 &
  13:19:00 &
  13:42:53 &
  13:31:41 &
  1433 &
  M3.4 &
  $\checkmark$ &
  37.2$\genfrac{}{}{0pt}{2}{+1.7}{-1.5}$ &
  27.6$\genfrac{}{}{0pt}{2}{+1.1}{-1.0}$ &
  46.7$\genfrac{}{}{0pt}{2}{+3.3}{-2.9}$ &
  19.1$\genfrac{}{}{0pt}{2}{+3.4}{-3.0}$ \\
57 &
  2014-08-25 &
  20:06:00 &
  20:35:56 &
  20:20:50 &
  1796 &
  M3.9 &
  $\checkmark$ &
  28$\genfrac{}{}{0pt}{2}{+0.8}{-0.7}$ &
  20.8$\genfrac{}{}{0pt}{2}{+0.5}{-0.5}$ &
  35.3$\genfrac{}{}{0pt}{2}{+1.4}{-1.3}$ &
  14.5$\genfrac{}{}{0pt}{2}{+1.5}{-1.4}$ \\
58 &
  2014-09-03 &
  13:20:00 &
  14:27:55 &
  13:54:11 &
  4075 &
  M2.5 &
  $\checkmark$ &
  37.2$\genfrac{}{}{0pt}{2}{+0.6}{-0.6}$ &
  49.1$\genfrac{}{}{0pt}{2}{+1.2}{-1.2}$ &
  25.2$\genfrac{}{}{0pt}{2}{+0.3}{-0.3}$ &
  -23.8$\genfrac{}{}{0pt}{2}{+1.3}{-1.2}$ \\
  
  \hline
\end{tabular}
\end{table}

\begin{table}
\contcaption{} 
\label{tab:continued}
\begin{tabular}{llllllccllll}
\hline\hline
\textbf{Num} &
  \textbf{Date} &
  \textbf{T$_{start}$} &
  \textbf{T$_{end}$} &
  \textbf{T$_{peak}$} &
  \textbf{Duration} &
  \textbf{GOES} &
  \textbf{CME} &
  \textbf{Period$_{Avg}$} &
  \textbf{Period$_{Impulsive}$} &
  \textbf{Period$_{Decay}$} &
  \textbf{Period drift} \\

  \textbf{} &
  \textbf{} &
  \textbf{} &
  \textbf{} &
  \textbf{} &
  (seconds) &
  \textbf{class} &
  \textbf{} &
  (seconds) &
  (seconds) &
  (seconds) &
  (seconds) \\
  \hline

59 &
  2014-09-10 &
  17:21:00 &
  18:08:55 &
  17:45:10 &
  2875 &
  X1.6 &
  $\checkmark$ &
  56.7$\genfrac{}{}{0pt}{2}{+2.2}{-2.0}$ &
  36.5$\genfrac{}{}{0pt}{2}{+1.0}{-0.9}$ &
  77$\genfrac{}{}{0pt}{2}{+4.4}{-3.9}$ &
  40.5$\genfrac{}{}{0pt}{2}{+4.5}{-4.0}$ \\
60 &
  2014-10-09 &
  01:30:00 &
  01:55:55 &
  01:43:23 &
  1555 &
  M1.3 &
   &
  59.7$\genfrac{}{}{0pt}{2}{+7.2}{-5.5}$ &
  98.5$\genfrac{}{}{0pt}{2}{+14.3}{-11.1}$ &
  20.9$\genfrac{}{}{0pt}{2}{+0.6}{-0.5}$ &
  -77.6$\genfrac{}{}{0pt}{2}{+14.3}{-11.1}$ \\
61 &
  2014-11-03 &
  11:23:00 &
  12:22:54 &
  11:53:30 &
  3594 &
  M2.2 &
  $\checkmark$ &
  44.1$\genfrac{}{}{0pt}{2}{+1.0}{-0.9}$ &
  31$\genfrac{}{}{0pt}{2}{+0.5}{-0.5}$ &
  57.2$\genfrac{}{}{0pt}{2}{+1.9}{-1.8}$ &
  26.2$\genfrac{}{}{0pt}{2}{+2.0}{-1.8}$ \\
62 &
  2014-11-04 &
  07:59:00 &
  09:16:53 &
  08:38:41 &
  4673 &
  M2.6 &
  $\checkmark$ &
  28.7$\genfrac{}{}{0pt}{2}{+0.3}{-0.3}$ &
  23.6$\genfrac{}{}{0pt}{2}{+0.2}{-0.2}$ &
  33.8$\genfrac{}{}{0pt}{2}{+0.5}{-0.5}$ &
  10.2$\genfrac{}{}{0pt}{2}{+0.6}{-0.5}$ \\
63 &
  2014-11-05 &
  18:50:00 &
  20:37:54 &
  19:44:38 &
  6474 &
  M2.9 &
  $\checkmark$ &
  56.1$\genfrac{}{}{0pt}{2}{+1.1}{-1.0}$ &
  30.3$\genfrac{}{}{0pt}{2}{+0.3}{-0.3}$ &
  81.8$\genfrac{}{}{0pt}{2}{+2.1}{-2.0}$ &
  51.4$\genfrac{}{}{0pt}{2}{+2.1}{-2.0}$ \\
64 &
  2014-11-06 &
  01:29:00 &
  01:48:54 &
  01:39:21 &
  1194 &
  M3.2 &
  $\checkmark$ &
  19$\genfrac{}{}{0pt}{2}{+0.5}{-0.4}$ &
  17.4$\genfrac{}{}{0pt}{2}{+0.5}{-0.5}$ &
  20.7$\genfrac{}{}{0pt}{2}{+0.7}{-0.7}$ &
  3.2$\genfrac{}{}{0pt}{2}{+0.9}{-0.9}$ $\dagger$ \\
65 &
  2014-11-06 &
  21:53:00 &
  22:38:49 &
  22:16:01 &
  2749 &
  M2.5 &
   &
  23.6$\genfrac{}{}{0pt}{2}{+0.3}{-0.3}$ &
  18.3$\genfrac{}{}{0pt}{2}{+0.2}{-0.2}$ &
  28.8$\genfrac{}{}{0pt}{2}{+0.6}{-0.6}$ &
  10.5$\genfrac{}{}{0pt}{2}{+0.7}{-0.6}$ \\
66 &
  2014-11-07 &
  10:13:00 &
  10:30:56 &
  10:22:15 &
  1076 &
  M1.0 &
   &
  17.1$\genfrac{}{}{0pt}{2}{+0.4}{-0.4}$ &
  14.7$\genfrac{}{}{0pt}{2}{+0.4}{-0.4}$ &
  19.5$\genfrac{}{}{0pt}{2}{+0.7}{-0.7}$ &
  4.8$\genfrac{}{}{0pt}{2}{+0.8}{-0.8}$ \\
67 &
  2014-11-15 &
  11:40:00 &
  12:25:55 &
  12:03:21 &
  2755 &
  M3.2 &
  $\checkmark$ &
  28.7$\genfrac{}{}{0pt}{2}{+0.4}{-0.4}$ &
  25.9$\genfrac{}{}{0pt}{2}{+0.5}{-0.5}$ &
  31.4$\genfrac{}{}{0pt}{2}{+0.7}{-0.7}$ &
  5.5$\genfrac{}{}{0pt}{2}{+0.9}{-0.8}$ \\
68 &
  2014-12-17 &
  04:25:00 &
  05:16:55 &
  04:50:06 &
  3115 &
  M8.7 &
  $\checkmark$ &
  64$\genfrac{}{}{0pt}{2}{+2.0}{-1.8}$ &
  56.8$\genfrac{}{}{0pt}{2}{+2.1}{-2.0}$ &
  71.2$\genfrac{}{}{0pt}{2}{+3.4}{-3.1}$ &
  14.3$\genfrac{}{}{0pt}{2}{+4.0}{-3.7}$ \\
69 &
  2014-12-18 &
  21:41:00 &
  22:14:55 &
  21:58:03 &
  2035 &
  M6.9 &
  $\checkmark$ &
  29$\genfrac{}{}{0pt}{2}{+0.7}{-0.6}$ &
  22.8$\genfrac{}{}{0pt}{2}{+0.5}{-0.5}$ &
  35.2$\genfrac{}{}{0pt}{2}{+1.3}{-1.2}$ &
  12.4$\genfrac{}{}{0pt}{2}{+1.4}{-1.3}$ \\
70 &
  2014-12-20 &
  00:11:00 &
  00:44:55 &
  00:28:00 &
  2035 &
  X1.8 &
  $\checkmark$ &
  21.3$\genfrac{}{}{0pt}{2}{+0.5}{-0.5}$ &
  10.6$\genfrac{}{}{0pt}{2}{+0.1}{-0.1}$ &
  31.9$\genfrac{}{}{0pt}{2}{+1.0}{-1.0}$ &
  21.2$\genfrac{}{}{0pt}{2}{+1.0}{-1.0}$ \\
71 &
  2015-03-02 &
  15:10:00 &
  15:45:56 &
  15:28:16 &
  2156 &
  M3.7 &
  $\checkmark$ &
  41.1$\genfrac{}{}{0pt}{2}{+1.3}{-1.2}$ &
  48.9$\genfrac{}{}{0pt}{2}{+2.3}{-2.1}$ &
  33.3$\genfrac{}{}{0pt}{2}{+1.1}{-1.0}$ &
  -15.6$\genfrac{}{}{0pt}{2}{+2.6}{-2.3}$ \\
72 &
  2015-03-06 &
  04:14:00 &
  05:39:54 &
  04:57:36 &
  5154 &
  M3.0 &
  $\checkmark$ &
  82.3$\genfrac{}{}{0pt}{2}{+2.1}{-2.0}$ &
  65.8$\genfrac{}{}{0pt}{2}{+1.7}{-1.6}$ &
  98.7$\genfrac{}{}{0pt}{2}{+3.9}{-3.6}$ &
  32.9$\genfrac{}{}{0pt}{2}{+4.3}{-4.0}$ \\
73 &
  2015-03-09 &
  14:22:00 &
  14:43:57 &
  14:33:45 &
  1317 &
  M4.5 &
  $\checkmark$ &
  11.9$\genfrac{}{}{0pt}{2}{+0.2}{-0.2}$ &
  14.1$\genfrac{}{}{0pt}{2}{+0.3}{-0.3}$ &
  9.8$\genfrac{}{}{0pt}{2}{+0.1}{-0.1}$ &
  -4.4$\genfrac{}{}{0pt}{2}{+0.3}{-0.3}$ \\
74 &
  2015-03-09 &
  23:29:00 &
  00:16:54 * &
  23:54:17 &
  2874 &
  M5.8 &
  $\checkmark$ &
  51.1$\genfrac{}{}{0pt}{2}{+1.7}{-1.5}$ &
  36.3$\genfrac{}{}{0pt}{2}{+0.9}{-0.9}$ &
  66$\genfrac{}{}{0pt}{2}{+3.2}{-2.9}$ &
  29.7$\genfrac{}{}{0pt}{2}{+3.3}{-3.0}$ \\
75 &
  2015-03-11 &
  16:11:00 &
  16:32:54 &
  16:21:38 &
  1314 &
  X2.1 &
  $\checkmark$ &
  22.3$\genfrac{}{}{0pt}{2}{+0.8}{-0.7}$ &
  13.3$\genfrac{}{}{0pt}{2}{+0.3}{-0.3}$ &
  31.3$\genfrac{}{}{0pt}{2}{+1.6}{-1.4}$ &
  18$\genfrac{}{}{0pt}{2}{+1.6}{-1.4}$ \\
76 &
  2015-03-12 &
  04:41:00 &
  04:50:53 &
  04:46:00 &
  593 &
  M3.2 &
   &
  11.4$\genfrac{}{}{0pt}{2}{+0.4}{-0.3}$ &
  8.4$\genfrac{}{}{0pt}{2}{+0.2}{-0.2}$ &
  14.3$\genfrac{}{}{0pt}{2}{+0.7}{-0.7}$ &
  5.9$\genfrac{}{}{0pt}{2}{+0.8}{-0.7}$ \\
77 &
  2015-03-12 &
  11:38:00 &
  12:01:53 &
  11:50:26 &
  1433 &
  M1.6 &
   &
  21.3$\genfrac{}{}{0pt}{2}{+0.5}{-0.5}$ &
  15.9$\genfrac{}{}{0pt}{2}{+0.4}{-0.3}$ &
  26.7$\genfrac{}{}{0pt}{2}{+1.0}{-1.0}$ &
  10.8$\genfrac{}{}{0pt}{2}{+1.1}{-1.0}$ \\
78 &
  2015-03-12 &
  12:09:00 &
  12:18:53 &
  12:14:23 &
  593 &
  M1.4 &
   &
  17.1$\genfrac{}{}{0pt}{2}{+0.9}{-0.8}$ &
  21.1$\genfrac{}{}{0pt}{2}{+1.6}{-1.4}$ &
  13.2$\genfrac{}{}{0pt}{2}{+0.6}{-0.6}$ &
  -7.9$\genfrac{}{}{0pt}{2}{+1.7}{-1.5}$ \\
79 &
  2015-03-12 &
  13:50:00 &
  14:25:53 &
  14:08:39 &
  2153 &
  M4.2 &
   &
  14.5$\genfrac{}{}{0pt}{2}{+0.1}{-0.1}$ &
  13.1$\genfrac{}{}{0pt}{2}{+0.2}{-0.2}$ &
  16$\genfrac{}{}{0pt}{2}{+0.2}{-0.2}$ &
  2.9$\genfrac{}{}{0pt}{2}{+0.3}{-0.3}$ $\dagger$ \\
80 &
  2015-03-13 &
  03:47:00 &
  04:14:54 &
  04:01:49 &
  1674 &
  M1.2 &
   &
  33.7$\genfrac{}{}{0pt}{2}{+1.3}{-1.2}$ &
  22.4$\genfrac{}{}{0pt}{2}{+0.6}{-0.6}$ &
  44.9$\genfrac{}{}{0pt}{2}{+2.5}{-2.3}$ &
  22.6$\genfrac{}{}{0pt}{2}{+2.6}{-2.4}$ \\
81 &
  2015-03-16 &
  10:39:00 &
  11:16:53 &
  10:57:59 &
  2273 &
  M1.6 &
   &
  27$\genfrac{}{}{0pt}{2}{+0.6}{-0.6}$ &
  18.3$\genfrac{}{}{0pt}{2}{+0.3}{-0.3}$ &
  35.6$\genfrac{}{}{0pt}{2}{+1.2}{-1.1}$ &
  17.3$\genfrac{}{}{0pt}{2}{+1.2}{-1.1}$ \\
82 &
  2015-03-17 &
  22:49:00 &
  23:59:55 &
  23:34:48 &
  4255 &
  M1.0 &
  $\checkmark$ &
  14.8$\genfrac{}{}{0pt}{2}{+0.1}{-0.1}$ &
  8.7$\genfrac{}{}{0pt}{2}{+0.1}{-0.1}$ &
  20.9$\genfrac{}{}{0pt}{2}{+0.2}{-0.2}$ &
  12.1$\genfrac{}{}{0pt}{2}{+0.2}{-0.2}$ \\
83 &
  2015-04-21 &
  07:08:00 &
  07:33:54 &
  07:20:48 &
  1554 &
  M1.0 &
   &
  19.1$\genfrac{}{}{0pt}{2}{+0.4}{-0.4}$ &
  15.7$\genfrac{}{}{0pt}{2}{+0.3}{-0.3}$ &
  22.5$\genfrac{}{}{0pt}{2}{+0.7}{-0.6}$ &
  6.8$\genfrac{}{}{0pt}{2}{+0.7}{-0.7}$ \\
84 &
  2015-06-21 &
  02:04:00 &
  03:03:32 &
  02:36:26 &
  3572 &
  M2.7 &
  $\checkmark$ &
  86.9$\genfrac{}{}{0pt}{2}{+3.3}{-3.0}$ &
  76.4$\genfrac{}{}{0pt}{2}{+3.4}{-3.1}$ &
  97.4$\genfrac{}{}{0pt}{2}{+5.6}{-5.0}$ &
  21$\genfrac{}{}{0pt}{2}{+6.6}{-5.9}$ \\
85 &
  2015-06-21 &
  02:06:00 &
  03:05:51 &
  02:36:45 &
  3591 &
  M2.6 &
  $\checkmark$ &
  84.8$\genfrac{}{}{0pt}{2}{+3.1}{-2.8}$ &
  75.4$\genfrac{}{}{0pt}{2}{+3.3}{-3.0}$ &
  94.1$\genfrac{}{}{0pt}{2}{+5.2}{-4.7}$ &
  18.7$\genfrac{}{}{0pt}{2}{+6.2}{-5.6}$ \\
86 &
  2015-06-22 &
  17:39:00 &
  19:06:55 &
  18:24:42 &
  5275 &
  M6.5 &
  $\checkmark$ &
  20.2$\genfrac{}{}{0pt}{2}{+0.1}{-0.1}$ &
  21.2$\genfrac{}{}{0pt}{2}{+0.2}{-0.2}$ &
  19.2$\genfrac{}{}{0pt}{2}{+0.1}{-0.1}$ &
  -1.9$\genfrac{}{}{0pt}{2}{+0.2}{-0.2}$ $\dagger$ \\
87 &
  2015-09-20 &
  17:32:00 &
  18:33:53 &
  18:01:50 &
  3713 &
  M2.1 &
  $\checkmark$ &
  81.9$\genfrac{}{}{0pt}{2}{+2.8}{-2.6}$ &
  70.2$\genfrac{}{}{0pt}{2}{+2.8}{-2.6}$ &
  93.6$\genfrac{}{}{0pt}{2}{+5.0}{-4.5}$ &
  23.4$\genfrac{}{}{0pt}{2}{+5.7}{-5.2}$ \\
88 &
  2015-09-28 &
  07:27:00 &
  07:42:53 &
  07:34:42 &
  953 &
  M1.1 &
   &
  13.5$\genfrac{}{}{0pt}{2}{+0.3}{-0.3}$ &
  13.8$\genfrac{}{}{0pt}{2}{+0.4}{-0.4}$ &
  13.3$\genfrac{}{}{0pt}{2}{+0.4}{-0.4}$ &
  -0.5$\genfrac{}{}{0pt}{2}{+0.6}{-0.5}$ $\dagger$ \\
89 &
  2015-10-02 &
  17:08:00 &
  17:27:54 &
  17:18:27 &
  1194 &
  M1.0 &
   &
  13.1$\genfrac{}{}{0pt}{2}{+0.3}{-0.2}$ &
  9.5$\genfrac{}{}{0pt}{2}{+0.2}{-0.1}$ &
  16.7$\genfrac{}{}{0pt}{2}{+0.5}{-0.5}$ &
  7.3$\genfrac{}{}{0pt}{2}{+0.5}{-0.5}$ \\
90 &
  2015-10-15 &
  23:27:00 &
  23:34:55 &
  23:31:49 &
  475 &
  M1.1 &
   &
  11.2$\genfrac{}{}{0pt}{2}{+0.4}{-0.4}$ &
  11.8$\genfrac{}{}{0pt}{2}{+0.6}{-0.6}$ &
  10.7$\genfrac{}{}{0pt}{2}{+0.5}{-0.5}$ &
  -1$\genfrac{}{}{0pt}{2}{+0.8}{-0.7}$ $\dagger$ \\
91 &
  2015-10-16 &
  06:11:00 &
  06:20:53 &
  06:16:31 &
  593 &
  M1.1 &
   &
  18.7$\genfrac{}{}{0pt}{2}{+1.0}{-0.9}$ &
  23.1$\genfrac{}{}{0pt}{2}{+2.0}{-1.7}$ &
  14.2$\genfrac{}{}{0pt}{2}{+0.7}{-0.6}$ &
  -8.9$\genfrac{}{}{0pt}{2}{+2.1}{-1.8}$ \\
92 &
  2015-10-17 &
  20:09:00 &
  20:36:54 &
  20:22:58 &
  1674 &
  M1.1 &
   &
  18.7$\genfrac{}{}{0pt}{2}{+0.5}{-0.5}$ &
  9.3$\genfrac{}{}{0pt}{2}{+0.1}{-0.1}$ &
  28.1$\genfrac{}{}{0pt}{2}{+1.0}{-0.9}$ &
  18.7$\genfrac{}{}{0pt}{2}{+1.0}{-0.9}$ \\
93 &
  2015-11-04 &
  11:55:00 &
  12:10:56 &
  12:03:17 &
  956 &
  M2.5 &
  $\checkmark$ &
  13.1$\genfrac{}{}{0pt}{2}{+0.3}{-0.3}$ &
  10.8$\genfrac{}{}{0pt}{2}{+0.2}{-0.2}$ &
  15.4$\genfrac{}{}{0pt}{2}{+0.5}{-0.5}$ &
  4.6$\genfrac{}{}{0pt}{2}{+0.6}{-0.5}$ \\
94 &
  2015-12-21 &
  00:52:00 &
  01:13:57 &
  01:03:04 &
  1317 &
  M2.8 &
  $\checkmark$ &
  21.2$\genfrac{}{}{0pt}{2}{+0.6}{-0.6}$ &
  14.8$\genfrac{}{}{0pt}{2}{+0.3}{-0.3}$ &
  27.7$\genfrac{}{}{0pt}{2}{+1.2}{-1.1}$ &
  12.8$\genfrac{}{}{0pt}{2}{+1.3}{-1.2}$ \\
95 &
  2015-12-22 &
  03:15:00 &
  03:52:55 &
  03:34:19 &
  2275 &
  M1.6 &
  $\checkmark$ &
  32.3$\genfrac{}{}{0pt}{2}{+0.7}{-0.7}$ &
  27.1$\genfrac{}{}{0pt}{2}{+0.7}{-0.6}$ &
  37.4$\genfrac{}{}{0pt}{2}{+1.3}{-1.2}$ &
  10.4$\genfrac{}{}{0pt}{2}{+1.4}{-1.3}$ \\
96 &
  2015-12-23 &
  00:23:00 &
  00:56:54 &
  00:40:46 &
  2034 &
  M4.7 &
  $\checkmark$ &
  38.1$\genfrac{}{}{0pt}{2}{+1.2}{-1.1}$ &
  29.1$\genfrac{}{}{0pt}{2}{+0.9}{-0.8}$ &
  47.1$\genfrac{}{}{0pt}{2}{+2.3}{-2.1}$ &
  18$\genfrac{}{}{0pt}{2}{+2.4}{-2.2}$ \\
97 &
  2016-07-23 &
  05:00:00 &
  05:31:52 &
  05:16:43 &
  1912 &
  M7.6 &
  $\checkmark$ &
  20.9$\genfrac{}{}{0pt}{2}{+0.4}{-0.3}$ &
  17.3$\genfrac{}{}{0pt}{2}{+0.3}{-0.3}$ &
  24.6$\genfrac{}{}{0pt}{2}{+0.6}{-0.6}$ &
  7.3$\genfrac{}{}{0pt}{2}{+0.7}{-0.7}$ \\
98 &
  2017-04-02 &
  07:50:00 &
  08:13:56 &
  08:02:56 &
  1436 &
  M5.3 &
  $\checkmark$ &
  22.6$\genfrac{}{}{0pt}{2}{+0.5}{-0.5}$ &
  19.5$\genfrac{}{}{0pt}{2}{+0.5}{-0.5}$ &
  25.7$\genfrac{}{}{0pt}{2}{+1.0}{-0.9}$ &
  6.1$\genfrac{}{}{0pt}{2}{+1.1}{-1.0}$
   \\

\hline

\end{tabular}
\begin{tabular}{l}
* indicates that the flaring event took place over midnight, so the end time of the flare occurs on the subsequent day to the date indicated. \\
$\dagger$ indicates that the period drift is smaller in magnitude than 4.09~s (twice the data cadence) and therefore the QPP is deemed to exhibit no period drift \\    in this study.
\end{tabular}
\end{table}


\label{lastpage}
\end{document}